\documentclass[10pt,twocolumn,twoside]{article}
\usepackage[T1]{fontenc}
\usepackage{lmodern}
\usepackage[a4paper,left=1.6cm,right=1.6cm,top=2.2cm,bottom=2.2cm,columnsep=0.7cm,headsep=0.5cm]{geometry}
\usepackage{amsmath,amssymb,amsthm,mathtools,bm}
\usepackage{graphicx}
\usepackage{array}
\usepackage[usenames,dvipsnames]{xcolor}
\usepackage[super,sort&compress,comma]{natbib}
\usepackage[font=footnotesize,labelfont=sc,labelsep=period,justification=justified,singlelinecheck=true]{caption}
\usepackage{balance}
\usepackage{microtype}
\usepackage{titlesec}
\usepackage{lettrine}
\usepackage{fancyhdr}
\usepackage[hidelinks]{hyperref}
\renewcommand{\thesection}{\Roman{section}}

\titleformat{\section}{\centering\small\bfseries}{\thesection.}{0.5em}{\MakeUppercase}
\titleformat{\subsection}{\centering\normalsize\bfseries}{\Alph{subsection}.}{0.5em}{}
\titleformat{\subsubsection}[runin]{\normalsize\itshape}{\thesubsubsection.}{0.5em}{}[.]
\titleformat{\paragraph}[runin]{\normalsize\itshape}{}{0pt}{}
\titlespacing*{\section}{0pt}{12pt plus 2pt minus 2pt}{6pt plus 1pt}
\titlespacing*{\subsection}{0pt}{9pt plus 2pt minus 1pt}{4pt plus 1pt}
\titlespacing*{\paragraph}{\parindent}{4pt plus 1pt}{0.5em}

\fancypagestyle{plain}{\fancyhf{}\fancyfoot[C]{\footnotesize\thepage}}

\newtheoremstyle{classic}{6pt}{6pt}{\itshape}{\parindent}{\scshape}{.}{0.5em}{}

\theoremstyle{classic}
\newtheorem{theorem}{Theorem}
\newtheorem{proposition}{Proposition}
\newtheorem{lemma}{Lemma}
\newtheorem{corollary}{Corollary}

\newcommand{\q}{\mathbf q}
\newcommand{\Q}{\mathbf Q}
\newcommand{\Qt}{\widetilde{\mathbf Q}}
\newcommand{\I}{\mathbf I}
\newcommand{\n}{\mathbf n}
\newcommand{\Kq}{\mathcal K}
\newcommand{\E}{\mathcal E}
\newcommand{\A}{\mathbf A}

\begin{document}

\makeatletter
\twocolumn[
  \begin{@twocolumnfalse}
  \begin{center}
    {\Large\bfseries Critical corner angles and defect exchange in confined nematics\\[3pt] with splay-bend anisotropy\par}
    \vspace{10pt}
    {\normalsize\textsc{Prabakaran Rajamanickam}\par}
    \vspace{3pt}
    {\small\itshape Department of Mathematics \& Statistics, University of Strathclyde, Glasgow G1 1XQ, United Kingdom\par}
    {\small prabakaran.rajamanickam@strath.ac.uk\par}
    \vspace{4pt}
    {\small November 14, 2026\par}
    \vspace{10pt}
    \begin{minipage}{0.84\textwidth}\small\noindent\hspace{1.5em}In confined nematics, the director turning imposed by the walls is absorbed by defects, and we show that the corners exchange these defects with the liquid. A tangentially anchored corner of opening angle $\omega$ takes one of several types, labelled by the number $m$ of half-turns of the director relative to the radius; neighbouring types differ by a bound defect of index $\pm\frac12$, and logarithmic energies decide when the corner sheds or absorbs it. For equal elastic constants the critical angles are $72^\circ$, $120^\circ$, $216^\circ$ and $360^\circ$, with windows of bistability between them, explaining the hidden defect near $72^\circ$ vertices and selecting the equilibria of polygons and discs. We introduce unequal splay and bend constants through the cubic invariant $Q_{lk}\,\partial_lQ_{ij}\,\partial_kQ_{ij}$ in reduced Landau--de Gennes theory; bounding the order by its bulk value makes the problem well posed, without further elastic terms, whenever both Frank constants are positive. At first order only the splay corner feels the anisotropy; other corner and defect energies are even in it and follow exactly by quadrature. The anisotropy moves the angle above which a splay corner sheds a $+\frac12$ defect from $104^\circ$ to $136^\circ$ as $K_3/K_1$ rises from $0.54$ to $1.86$, and so predicts which polygons prefer splay corners and which interior defects. Finite-element computations for polygons of three to eight sides and the disc confirm this: the exchange occurs within the predicted range by migration of $+\frac12$ defects into the corners, near which states with zero, one and two splay corners are nearly degenerate.
    \end{minipage}\par
    \vspace{8pt}
    \rule{0.2\textwidth}{0.4pt}\par
  \end{center}
  \vspace{6pt}
  \end{@twocolumnfalse}
]
\makeatother
\thispagestyle{plain}

\section{Introduction}

\lettrine[lines=2,lhang=0.05,nindent=0.1em]{A}{\ nematic} liquid crystal held in a shallow well, between plates that impose degenerate planar anchoring and side walls
that impose tangential anchoring, usually has several stable equilibria. The walls force the director to turn as the
boundary is traversed; the turning must be accommodated by defects in the liquid or at the corners of the well; and the
different ways of accommodating it are separated by energy barriers. Such multistability was observed in square wells by
Tsakonas \emph{et al.}~\cite{Tsakonas07}, and has since been studied in squares, polygons and discs within the
Oseen--Frank and Landau--de Gennes theories~\cite{deGP93,MN14,LMR12,KM14,ML16,HMZ20,HYZMZ21,RHAM26}. When the layer is
thin, the order tensor may be taken to be a symmetric, traceless $2\times2$ tensor, and the Landau--de Gennes energy
reduces to the \emph{reduced} Landau--de Gennes (rLdG) energy~\cite{GMS15}.

In such wells the corners are where much of the frustration is resolved, and they have been studied from several sides.
The director field in a wedge can be constructed by conformal mapping~\cite{DM12}; on a sawtooth substrate, disclinations
of non-half-integer index sit at the wedges and apexes and give the surface free energy a contribution proportional to
$q\ln q$, $q$ being the wavenumber of the substrate~\cite{REPP10}; a flat wall with strong anchoring repels a disclination, and one with weak anchoring
may expel it~\cite{BS03}; and the equilibrium positions of defects in confined two-dimensional nematics have been
computed~\cite{TL23}. In polygonal wells the states are classified by the profiles of the director at the vertices, which
in the limit of a small well change from splay to bend at a vertex angle of $90^\circ$ and from bend to its reverse at
$270^\circ$~\cite{HMZ20,RHAM26}.  For walls with tangential anchoring, Yao, Zhang and Chen~\cite{Yao22} classified the textures at a corner as splay or
bend, for convex and reflex angles alike, derived from them the total index of the defects in a confined two-dimensional
nematic, and observed that a bend corner can be regarded as a splay corner carrying a \emph{hidden} defect of index
$-\frac12$; which texture a corner adopts, they noted, is not fixed by its angle but by the free energy of the whole
state. In an isosceles triangle the apex changes from a splay profile with a nearby defect of index $-\frac12$ to a bend
profile as the apex angle passes through about $72^\circ$~\cite{RHAM26}. What has been missing is the energetic rule that
decides, for a given angle, whether a corner gives up a defect to the liquid or takes one in.

We supply such a rule. In a domain large compared with the coherence length the energy of an equilibrium is dominated by
terms proportional to the logarithm of the size of the domain, one for each defect and one for each corner. A corner of
opening angle $\omega$ can take several types, distinguished by the number of half-turns of the director relative to the
radius as the corner is crossed, and neighbouring types differ by a bound defect of index $\pm\frac12$. Comparing the
logarithmic energies of a corner with and without such a defect gives the force on the defect, and so the critical angles
at which the corner sheds it into the liquid or absorbs it: $72^\circ$, $120^\circ$, $216^\circ$ and $360^\circ$ for a nematic
with a single elastic constant, with windows between them in which either type is stable. The hidden defect is the case
$72^\circ$. Applied corner by corner, the rule predicts which states the regular polygons select.

The polygons have fixed angles, so that in a nematic with a single elastic constant the rule can be tested only at a few
discrete points. Real nematics, however, have unequal splay and bend constants: in 5CB bend exceeds splay by about one
half~\cite{Bradshaw85}, and in bent-core nematics bend is the softer, $K_3/K_1$ being about one half~\cite{Sathyanarayana10}. We
show that the anisotropy acts on the corners almost entirely through one critical angle, and so provides a continuous
control that moves that angle across the fixed angles of the polygons. This is the second theme of the paper, and the means
by which the first is tested.

Introducing the anisotropy into the rLdG model requires care. It is usually introduced into Landau--de Gennes theory
through the quadratic invariant $|\nabla\!\cdot\!\Q|^2$~\cite{HHMZ22,han2022uniaxial,rajamanickam2025colloidal}, which, for a
uniaxial tensor of constant order, adds equally to the splay and bend constants and leaves the twist constant unchanged: it
distinguishes twist from splay and bend, but not splay from bend. In a thin layer under strong planar confinement twist is
absent, and this anisotropy has nothing to act on (\S\ref{sec:cubic}). The lowest invariant that distinguishes splay from
bend is cubic in $\Q$, $Q_{lk}\,\partial_lQ_{ij}\,\partial_kQ_{ij}$~\cite{ST83,BM84,MN14}, and with it the Landau--de Gennes energy is
unbounded below~\cite{BM10,Ball17,IXZ15}: at large order the cubic term gives one direction of distortion a negative stiffness.
Ball and Majumdar~\cite{BM10} replaced the Landau potential by a singular potential that keeps the eigenvalues of $\Q$ within
their microscopic range, a route used to compute disclinations in anisotropic nematics~\cite{Myers24}, and Golovaty, Novack and
Sternberg~\cite{GNS21} added elastic terms quartic in $\Q$. We take a third course, which adds nothing to the energy: we
require only that the order nowhere exceed its value in the bulk nematic, which in the one-constant model is a property of
the equilibria, and show that this is enough for every anisotropy at which both Frank constants are positive.

The paper is arranged as follows. The model is formulated in \S\ref{sec:form}, and its well-posedness established in
\S\ref{sec:bound}. In \S\ref{sec:defects} we derive the energies of corners and defects, the types of a corner and their
critical angles, the selection rules for polygons and the disc, and, in \S\ref{sec:exact}, the exact energies beyond first
order in the anisotropy. The numerical method is described in \S\ref{sec:numerics}, and the computations, which test the rules
in polygons of three to eight sides and in the disc, in \S\ref{sec:results}. The proofs, and the discrete form of the cubic
term, are given in the appendices.

\section{Formulation}\label{sec:form}

\subsection{The order tensor}\label{sec:order}

In a thin layer with degenerate planar anchoring on the plates, the orientational order is described by a symmetric,
traceless $2\times2$ tensor, which has two independent components~\cite{GMS15},
\begin{equation}\label{eq:Q}
\Q=\begin{pmatrix}Q_{11}&Q_{12}\\ Q_{12}&-Q_{11}\end{pmatrix}
=s\Big(\n\otimes\n-\tfrac12\I\Big),
\end{equation}
with $\n=(\cos\varphi,\sin\varphi)$. Here $s$ is the scalar order parameter and $\varphi$, defined modulo $\pi$, is the angle
of the director $\n$; a point at which $s=0$ is a defect.

Two order tensors must be distinguished. The \emph{microscopic} order tensor is the ensemble average
$\langle\mathbf m\otimes\mathbf m-\frac12\I\rangle$ over the orientations $\mathbf m$ of the molecules. Its eigenvalues $\pm\frac12S$
lie in $[-\frac12,\frac12]$, so that its order $S$ lies between $0$ and $1$, and $S=1$ corresponds to perfect alignment. The
tensor $\Q$ of Landau--de Gennes theory is related to this average but is not identical with it. It is the variable of a polynomial expansion of the free
energy about the isotropic state, truncated at fourth order, whose coefficients absorb the terms that have been left out
and the effect of the fluctuations; it is a coarse-grained, renormalised order parameter, while the microscopic tensor is a
bare one. The magnitude of $\Q$ is therefore fixed by the ratio of the Landau coefficients, through the bulk order $s_+$ of
\S\ref{sec:energy}; these coefficients may be calibrated against measurements, but a polynomial model does not of itself respect the
bound $S\le1$ of the microscopic tensor, and that bound need not carry over to it. What is most directly comparable with
the microscopic tensor, or with experiment, is the order relative to its bulk value: $\Q/s_+$ is to be compared with the microscopic tensor divided by its bulk order. It is in this
normalised form that the order enters the theory below.

\subsection{Why the anisotropy must be cubic}\label{sec:cubic}

The elastic energy density is a scalar built from $\Q$ and its gradient. Besides
$|\nabla\Q|^2=\partial_kQ_{ij}\,\partial_kQ_{ij}$, two invariants are quadratic in the gradient,
$\partial_jQ_{ij}\,\partial_kQ_{ik}=|\nabla\!\cdot\!\Q|^2$ and $\partial_jQ_{ik}\,\partial_kQ_{ij}$; their difference is a divergence, so
that its integral depends only on the values of $\Q$ on the boundary, and they have the same effect in the interior.
Consider a director field of uniform order, $s$ constant. Then
\begin{equation}
(\nabla\!\cdot\!\Q)_i=s\big[(\n\cdot\nabla)n_i+n_i\,\nabla\!\cdot\!\n\big],
\end{equation}
and, since $(\n\cdot\nabla)\n$ is perpendicular to $\n$,
\begin{equation}\label{eq:divQ}
|\nabla\!\cdot\!\Q|^2=s^2\big[(\nabla\!\cdot\!\n)^2+|(\n\cdot\nabla)\n|^2\big]=s^2|\nabla\varphi|^2 .
\end{equation}
The splay $(\nabla\!\cdot\!\n)^2$ and the bend $|(\n\cdot\nabla)\n|^2$ appear with equal weight, and the term merely adds
to the single elastic constant, as does $|\nabla\Q|^2=2s^2|\nabla\varphi|^2$. In three dimensions the same calculation gives
$K_1=K_3\neq K_2$: the quadratic invariants separate twist from splay and bend, and it is only because twist is absent in
the plane that they leave a single constant here. In two dimensions no quadratic invariant distinguishes splay from bend
where the order is uniform; they do so only through the gradients of $s$, in the defect cores and the boundary layers.

A term that distinguishes them must contain $\Q$ itself, so that the gradient can be resolved along and across the
director. The lowest such term is cubic~\cite{ST83,BM84},
\begin{equation}\label{eq:cubic}
Q_{lk}\,\partial_lQ_{ij}\,\partial_kQ_{ij}.
\end{equation}
For uniform order, $\partial_lQ_{ij}\,\partial_kQ_{ij}=2s^2\,\partial_l\varphi\,\partial_k\varphi$, and contraction with
$Q_{lk}=s(n_ln_k-\frac12\delta_{lk})$ gives
\begin{align}\label{eq:cubicOF}
Q_{lk}\,\partial_lQ_{ij}\,\partial_kQ_{ij}&=s^3\big[(\n\cdot\nabla\varphi)^2-(\n^\perp\!\cdot\nabla\varphi)^2\big]\nonumber\\
&=s^3\big[|(\n\cdot\nabla)\n|^2-(\nabla\!\cdot\!\n)^2\big],
\end{align}
with $\n^\perp=(-\sin\varphi,\cos\varphi)$. The rate at which the director turns as one moves along it is the bend, and the
rate at which it turns as one moves across it is the splay; the cubic term therefore adds to the cost of bend what it
removes from the cost of splay. Since twist and saddle-splay are absent in two dimensions, this one term reproduces the
Oseen--Frank energy in full; other cubic invariants differ from it only in terms that involve $\nabla s$.

The term is odd in $\Q$. Changing $\Q$ into $-\Q$ rotates the director through $\frac12\pi$ everywhere, which exchanges
splay and bend, and the cubic term changes sign, as it must.

\subsection{The energy}\label{sec:energy}

The free energy of the layer, per unit thickness, is
\begin{multline}\label{eq:Fdim}
\mathcal F[\Q]=\int_\Omega\Big[\frac{L_1}2|\nabla\Q|^2+\frac{L_3}2\,Q_{lk}\,\partial_lQ_{ij}\,\partial_kQ_{ij}\\
+\frac A2\operatorname{tr}\Q^2+\frac C4(\operatorname{tr}\Q^2)^2\Big]dA
+\oint_{\partial\Omega}\frac W2|\Q-\Q_b|^2\,ds,
\end{multline}
the sum of the elastic energy, the bulk potential, with $A<0$ in the nematic phase, and a Rapini--Papoular anchoring
energy on the side walls, of strength $W$. The preferred tensor on the wall is
$\Q_b=s_+(\boldsymbol\tau\otimes\boldsymbol\tau-\frac12\I)$, where $\boldsymbol\tau$ is the unit tangent to the wall and
$s_+=(-2A/C)^{1/2}$ is the order of the uniform nematic, the minimiser of the bulk potential. We write $L_1$, $L_2$ and $L_3$ for the coefficients of
$|\nabla\Q|^2$, $|\nabla\!\cdot\!\Q|^2$ and the cubic invariant \eqref{eq:cubic}. The remaining quadratic invariant,
$\partial_jQ_{ik}\,\partial_kQ_{ij}$, to which the label $L_3$ is often given, differs from $|\nabla\!\cdot\!\Q|^2$ by a divergence
(\S\ref{sec:cubic}), and so contributes only a boundary term; such terms are of relative size $l_{\rm ex}/R=1/(\gamma\lambda)$
compared with the anchoring energy, and are neglected, as is usual when the anchoring is strong. The $L_2$ term only adds to
$L_1$ in two dimensions (\S\ref{sec:cubic}) and is omitted. The cubic coefficient is called $L_4$ in Refs.~\cite{BM10,IXZ15}
and $L_6$ in Ref.~\cite{MN14}. Of the cubic invariants~\cite{ST83,BM84}, \eqref{eq:cubic} is the
one retained in Ref.~\cite{MN14}, and, as \S\ref{sec:cubic} shows, the only one needed in two dimensions.

We measure lengths in units of the circumradius $R$ of the domain, $\Q$ in units of $s_+$, and energies in units of
$L_1s_+^2$, and represent the order tensor by the vector $\q=(q_1,q_2)$ of its components,
\begin{gather}
\Q=s_+\Qt(\q),\qquad \Qt(\q)=\begin{pmatrix}q_1&q_2\\ q_2&-q_1\end{pmatrix},\nonumber\\
\q=\frac{s}{2s_+}(\cos2\varphi,\sin2\varphi).\label{eq:qdef}
\end{gather}
The length $|\q|$ is half the order relative to the bulk, the eigenvalues of $\Qt$ are $\pm|\q|$, and the uniform nematic
has $|\q|=\frac12$; by \S\ref{sec:order}, $\q$ is the physically meaningful, normalised measure of the order. With
$|\nabla\Q|^2=2s_+^2|\nabla\q|^2$ and $\partial_lQ_{ij}\,\partial_kQ_{ij}=2s_+^2\,\partial_l\q\cdot\partial_k\q$,
and the energy measured from that of the uniform nematic, \eqref{eq:Fdim} becomes
\begin{align}\label{eq:F}
F[\q]&=\int_\Omega\Big[\sum_{a=1}^2\nabla q_a\cdot\A(\q)\nabla q_a+2\lambda^2\big(|\q|^2-\tfrac14\big)^2\Big]dA \nonumber\\
&\quad +\gamma\lambda\oint_{\partial\Omega}|\q-\q_b|^2\,ds,
\end{align}
in which $\q_b$ represents $\Q_b$ as $\q$ represents $\Q$, and
\begin{equation}\label{eq:Amat}
\A(\q)=\I+\eta\,\Qt(\q)=\begin{pmatrix}1+\eta q_1&\eta q_2\\ \eta q_2&1-\eta q_1\end{pmatrix}.
\end{equation}
Written out, the elastic density is
\begin{equation}\label{eq:elastic}
|\nabla\q|^2+\eta\big[q_1\big(|\partial_x\q|^2-|\partial_y\q|^2\big)+2q_2\,\partial_x\q\cdot\partial_y\q\big].
\end{equation}
Three parameters appear,
\begin{gather}
\lambda=\frac R\xi,\qquad\gamma=\frac{\xi}{l_{\rm ex}},\qquad\eta=\frac{L_3s_+}{L_1},\label{eq:params}\\
\text{where} \quad \xi=\left(\frac{L_1}{|A|}\right)^{1/2},\qquad l_{\rm ex}=\frac{L_1}W .\nonumber
\end{gather}
The first is the size of the domain in units of the coherence length $\xi$, the radius of a defect core; the second is the
ratio of $\xi$ to the extrapolation length $l_{\rm ex}$; and the third is the anisotropy. In practice $\xi$ is of order
1--10\,nm and $l_{\rm ex}$ is at least 10\,nm, so that $\gamma\lesssim1$~\cite{Raj26a,Raj26b}. We take $\gamma=1$, the
strongest anchoring that can be realised, and $\lambda=100$, a well of radius about a micrometre. The anchoring is then
strong, $\gamma\lambda=R/l_{\rm ex}=100$; strong anchoring is a consequence of the size of the domain, $R\gg l_{\rm ex}$, and not
of a large anchoring strength. The corrections to the strong-anchoring limit, and the surface defects that appear when it
fails, are analysed in Refs.~\cite{Raj26a,Raj26b}.

The domains of interest are therefore large compared with the coherence length, and we do not consider small ones,
$\lambda$ of order ten or less, for two reasons. First, Landau--de Gennes theory is a mean-field theory of a coarse-grained
order parameter (\S\ref{sec:order}), and it ceases to be reliable when the domain is only a few coherence lengths across,
for the scale of the coarse-graining is then comparable with the domain itself. Secondly, since $\gamma\lesssim1$ the
dimensionless anchoring strength satisfies $\gamma\lambda\lesssim\lambda$, so that a small domain is necessarily weakly anchored; the walls
can then no longer sustain the order, and the nematic is expected to lose it and become isotropic~\cite{Raj26a,Raj26b}. The
ordered equilibria that are found as $\lambda\to0$ with the director prescribed on the walls presuppose an anchoring energy
that grows without bound as the domain shrinks, which no real wall provides, and they lie outside the range of the
physical problem. We take $\lambda=100$ throughout, and have checked the behaviour of the bound of \S\ref{sec:bound} at
$\lambda=20$ and $30$, with $\gamma=1$ and $0.3$ (\S\ref{sec:boundres}).

The matrix $\A(\q)$ plays the part of the conductivity in a problem of heat conduction: the cubic term makes the elastic
medium anisotropic, with principal axes along and across the local director and an anisotropy proportional to the local
order. Its eigenvalues are $1\pm\eta|\q|$.

\subsection{Splay and bend}

The meaning of $\eta$ is found by evaluating \eqref{eq:F} for a director field of bulk order,
$\q=\frac12(\cos2\varphi,\sin2\varphi)$. Then $\partial_l\q\cdot\partial_k\q=\partial_l\varphi\,\partial_k\varphi$ and $\Qt=\n\otimes\n-\frac12\I$, and the
elastic density \eqref{eq:elastic} becomes
\begin{equation}\label{eq:OF}
\Big(1-\frac\eta2\Big)(\nabla\!\cdot\!\n)^2+\Big(1+\frac\eta2\Big)|(\n\cdot\nabla)\n|^2,
\end{equation}
the Oseen--Frank energy with
\begin{gather}
K_1=2L_1s_+^2-L_3s_+^3,\qquad K_3=2L_1s_+^2+L_3s_+^3,\label{eq:K}\\
\frac{K_3}{K_1}=\frac{2+\eta}{2-\eta},\qquad \eta=2\,\frac{K_3-K_1}{K_3+K_1}.\nonumber
\end{gather}
Positive $\eta$ makes bend the more expensive distortion, negative $\eta$ splay; both $K_1$ and $K_3$ are positive if and
only if $|\eta|<2$. Because the anisotropic part varies as $s^3$ and the isotropic part as $s^2$, the anisotropy fades in
the defect cores, where the order falls to zero. Representative values are given in Table~\ref{tab:materials}; the
computations of \S\ref{sec:results} span $-0.6\le\eta\le0.6$, that is, $0.54\le K_3/K_1\le1.86$.

\begin{table}[h]
\small
  \caption{\ Representative anisotropies; the ratios are approximate and depend on temperature}
  \label{tab:materials}
  \begin{tabular*}{\columnwidth}{@{\extracolsep{\fill}}lcc}
    \hline
    Nematic & $K_3/K_1$ & $\eta$ \\
    \hline
    5CB, room temperature\cite{Bradshaw85} & $\approx1.5$ & $\approx0.4$ \\
    One-constant model & $1$ & $0$ \\
    Bent-core nematic\cite{Sathyanarayana10} & $\approx0.5$ & $\approx-0.6$ \\
    \hline
  \end{tabular*}
\end{table}

\subsection{The domains and the anchoring}

The domain $\Omega$ is either the regular $N$-gon inscribed in the unit circle, with one edge horizontal, or the unit disc.
On the boundary $\q_b=\frac12(\cos2\theta_\tau,\sin2\theta_\tau)$, $\theta_\tau$ being the angle of the tangent. On a polygon $\q_b$ is
constant on each edge and jumps at the corners. Because the anchoring enters through the surface energy in \eqref{eq:F},
rather than as a condition imposed on $\q$, $\q_b$ need only be bounded on $\partial\Omega$, and these jumps require no special
treatment.

\subsection{Symmetries}

A rotation of the domain through an angle $\alpha$ rotates the director through $\alpha$, and so acts on $\q$, which depends
on $2\varphi$, as a rotation of the $\q$-plane through $2\alpha$; a reflection in a line at angle $\beta$ acts as the reflection
of the $\q$-plane in the line at angle $2\beta$. The energy \eqref{eq:F} is invariant under the dihedral group $D_N$ of the
polygon, or the orthogonal group of the disc, acting in this way, so that every equilibrium that breaks a symmetry has
copies of equal energy.

The energy is also invariant under
\begin{equation}\label{eq:homeo}
(\q,\;\q_b,\;\eta)\;\longmapsto\;(-\q,\;-\q_b,\;-\eta).
\end{equation}
Changing $\q_b$ into $-\q_b$ replaces tangential by homeotropic anchoring. Homeotropic anchoring with anisotropy $\eta$ is
therefore tangential anchoring with anisotropy $-\eta$, and it suffices to study tangential anchoring for both signs of
$\eta$. In the one-constant model the two anchorings are equivalent; with $\eta\neq0$ they are not.

\section{The bound on the order}\label{sec:bound}

\subsection{Negative stiffness}\label{sec:negstiff}

At a point where the order is $|\q|$, the elastic density \eqref{eq:elastic} is a quadratic form in the gradient of $\q$,
with matrix $\A(\q)$ for each component. Its eigenvalues are $1+\eta|\q|$ and $1-\eta|\q|$; the second is negative when
$|\eta||\q|>1$, and then a distortion whose gradient lies along the corresponding eigenvector has negative stiffness.
A field that oscillates in that direction, with wavenumber $k$ and amplitude $\delta$, lowers the elastic energy by an
amount proportional to $\delta^2k^2$, which grows without limit as $k$ increases, while the Landau and anchoring energies
remain bounded. The quartic Landau potential cannot prevent this, for it does not involve the gradient $(k=0)$. Among fields of
finite energy, $F$ is therefore unbounded below for every $\eta\neq0$~\cite{BM10,IXZ15}.

\subsection{The bound}\label{sec:thebound}

The instability requires an order larger than $1/|\eta|$, which for $|\eta|<2$ exceeds the bulk value $\frac12$. It is
therefore removed by any bound on the order that keeps $|\eta||\q|$ below one, and nothing more is needed. The natural bound
in Landau--de Gennes theory is that the order nowhere exceeds its value in the bulk nematic,
\begin{equation}\label{eq:bound}
|\q|\le\tfrac12 ,
\end{equation}
for, by \S\ref{sec:order}, it is the order relative to its bulk value that is most directly physical. The microscopic bound
$S\le1$ does not provide an alternative. Transferred to $\Q$ it would read $|\q|\le1/(2s_+)$; but $s_+$ depends on the
truncation of the Landau expansion and on how its coefficients are calibrated, so that such a bound would tie the anisotropy
that the theory can represent to the details of the polynomial model rather than to the material, and it would permit the order to exceed its bulk value, which nothing
in the problem favours.

In the one-constant model, $\eta=0$, the bound \eqref{eq:bound} is not an assumption but a property of the equilibria.
These satisfy $\nabla^2\q=\lambda^2(4|\q|^2-1)\q$, so that
\begin{equation}
\tfrac12\nabla^2|\q|^2=|\nabla\q|^2+\lambda^2(4|\q|^2-1)|\q|^2,
\end{equation}
which is positive wherever $|\q|>\frac12$. By the maximum principle $|\q|^2$ can then have no interior maximum where it
exceeds $\frac14$; nor can it have one on the boundary, where the anchoring condition gives
$\frac12\partial_\nu|\q|^2=-\gamma\lambda(|\q|^2-\q\cdot\q_b)$, which is negative when $|\q|>\frac12=|\q_b|$. The same conclusion
follows from a variational argument that we shall need again. Let $P$ be the radial projection of the $\q$-plane onto
the disc of radius $\frac12$,
\begin{equation}\label{eq:proj}
P(\q)=\begin{cases}\q, & |\q|\le\frac12,\\[2pt] \dfrac{\q}{2|\q|}, & |\q|>\frac12,\end{cases}
\end{equation}
which reduces the order to the bulk value wherever it exceeds it and leaves the director unchanged. Replacing a field
$\q(\mathbf x)$ by $P(\q(\mathbf x))$ lowers the bulk potential wherever the order is reduced; it does not increase the
Dirichlet energy, because $P$ does not increase distances in the $\q$-plane; and it does not increase the anchoring energy,
because $\q_b$ lies in the disc and $P$ moves every point towards it. A minimiser of the one-constant energy therefore
satisfies \eqref{eq:bound}. Physically, nothing in the problem favours an order above the bulk value: there are no
applied fields, and the walls prefer the bulk order. With $\eta\neq0$ the argument fails, because the conductivity
$\A(\q)$ changes under the projection, and \eqref{eq:bound} becomes a constraint. For the proofs we write the result for a
general bound $\bar q$: the admissible fields are
\begin{equation}
\Kq_{\bar q}=\big\{\q\in H^1(\Omega;\mathbb R^2):\ |\q(\mathbf x)|\le\bar q\ \text{almost everywhere}\big\},
\end{equation}
$H^1$ denoting the fields of finite Dirichlet energy, and the problem is to minimise $F$ over $\Kq_{\bar q}$, with
$\bar q=\frac12$. No term is added to the energy.

\subsection{Existence of minimisers}

\begin{theorem}\label{thm:main}
Let $\Omega$ be a bounded domain with a Lipschitz boundary (as polygons and discs are), let $\q_b$ be bounded on $\partial\Omega$,
$\lambda>0$, $\gamma\ge0$ and $\bar q>0$.
\begin{enumerate}
\item[(i)] If $|\eta|\,\bar q<1$, then $F\ge0$ on $\Kq_{\bar q}$, and $F$ attains its minimum on $\Kq_{\bar q}$.
\item[(ii)] If $|\eta|\,\bar q>1$, then $F$ is unbounded below on $\Kq_{\bar q}$.
\end{enumerate}
\end{theorem}

The proof is given in Appendix~\ref{app:proof}; its physical content is simple. Under the bound the eigenvalues of
$\A(\q)$ are at least $1-|\eta|\bar q>0$, so that the elastic energy controls the gradient; the order is bounded; and a
sequence of fields whose energies approach the least value has a convergent subsequence. The energy is convex in the
gradient, which ensures that the limit has no more energy than the sequence. If instead $|\eta|\bar q>1$, a uniform field
of order between $1/|\eta|$ and $\bar q$ has a direction of negative stiffness, and the oscillation described above lowers the
energy without limit while respecting the bound. The threshold $|\eta|\bar q=1$ is therefore sharp.

\begin{corollary}\label{cor:range}
With the bound \eqref{eq:bound}, $\bar q=\frac12$, the problem is well posed if and only if, up to the borderline case,
$|\eta|<2$; that is, exactly when both Frank constants are positive.
\end{corollary}

The bound therefore removes nothing that the theory could represent: the anisotropies for which the constrained problem is
well posed are exactly those for which the Oseen--Frank energy \eqref{eq:OF} is itself positive.

\subsection{The equilibrium equations}

Because $\Kq_{\bar q}$ is convex, a minimiser $\q$ cannot lower its energy by moving towards any other admissible field
$\mathbf v$: $\frac{d}{dt}F[\q+t(\mathbf v-\q)]\big|_{t=0^+}\ge0$. This variational inequality, written out in
Appendix~\ref{app:proof}, is the precise form of the equilibrium conditions. Where the bound is not attained it reduces to
the Euler--Lagrange equations. Formally, with a multiplier $\mu\ge0$ that vanishes except where $|\q|=\bar q$, they read
\begin{multline}\label{eq:EL}
\nabla\!\cdot\!\big[\A(\q)\nabla\q\big]=\frac\eta2\begin{pmatrix}|\partial_x\q|^2-|\partial_y\q|^2\\[2pt]
2\,\partial_x\q\cdot\partial_y\q\end{pmatrix}\\
+4\lambda^2\big(|\q|^2-\tfrac14\big)\q+\mu\,\q\quad\text{in }\Omega,
\end{multline}
with the boundary condition
\begin{equation}\label{eq:BC}
\boldsymbol\nu\cdot\A(\q)\nabla\q=-\gamma\lambda\,(\q-\q_b)\quad\text{on }\partial\Omega,
\end{equation}
$\boldsymbol\nu$ being the outward normal. Here $\A(\q)\nabla\q$ is the flux whose $a$th component is the vector $\A(\q)\nabla q_a$,
and the divergence and the product with $\boldsymbol\nu$ act on its spatial index, leaving two-component vectors:
$\big(\nabla\!\cdot\![\A\nabla\q]\big)_a=\partial_l(A_{lk}\,\partial_kq_a)$ and $(\boldsymbol\nu\cdot\A\nabla\q)_a=\nu_lA_{lk}\,\partial_kq_a$.
The left side of \eqref{eq:EL} is the divergence of an anisotropic flux, the elastic analogue of the heat flux in an
anisotropic conductor; the first term on the right arises because the conductivity itself depends on $\q$, and is the
derivative of the elastic density \eqref{eq:elastic} with respect to $\q$ at fixed gradient. With $\eta=0$ and $\mu=0$ the
equations are those of the one-constant model, $\nabla^2\q=\lambda^2(4|\q|^2-1)\q$ with $\partial_\nu\q=-\gamma\lambda(\q-\q_b)$.

Whether the bound is attained by the equilibria of interest decides how they are to be understood. If it is not, they
solve \eqref{eq:EL}--\eqref{eq:BC} with $\mu=0$, and the bound has served only to make the minimisation well posed. When it
can be attained is found next.

\subsection{When the order rises above its bulk value}\label{sec:rise}

In the one-constant model the order never exceeds its bulk value (\S\ref{sec:thebound}). With $\eta\neq0$ it may, and the reason, and the
anisotropy at which it first happens, can be found from the energy at a single point.

Write the order relative to its bulk value as $\tilde s=s/s_+=2|\q|$, so that $\q=\frac12\tilde s(\cos2\varphi,\sin2\varphi)$ and the
bound \eqref{eq:bound} reads $\tilde s\le1$. Away from the defect cores the order differs from its bulk value only by an amount of
order $\lambda^{-2}$~\cite{di2020landau,Raj26a,Raj26b}, so that its gradient may be neglected beside that of the director; this is the
approximation of \S\ref{sec:defects}, and it fails only within a few coherence lengths of a defect or a corner. With
$\nabla\q=\tilde s\,\mathbf e^\perp\otimes\nabla\varphi$, where $\mathbf e^\perp=(-\sin2\varphi,\cos2\varphi)$ is a unit vector,
\begin{equation}
|\nabla\q|^2=\tilde s^2\,|\nabla\varphi|^2,\qquad \partial_l\q\cdot\partial_k\q=\tilde s^2\,\partial_l\varphi\,\partial_k\varphi,
\end{equation}
and, since $\Qt(\q)=\tilde s\,(\n\otimes\n-\frac12\I)$, the cubic part of \eqref{eq:elastic} is
\begin{equation}
\eta\,\tilde s^3\Big[(\n\cdot\nabla\varphi)^2-\tfrac12|\nabla\varphi|^2\Big]=\frac\eta2\,\tilde s^3\big(B-S\big),
\end{equation}
where $S=(\n^\perp\!\cdot\nabla\varphi)^2$ and $B=(\n\cdot\nabla\varphi)^2$ are the splay and the bend, as in \eqref{eq:cubicOF}, and
$S+B=|\nabla\varphi|^2$. Let
\begin{equation}\label{eq:zeta}
G=S+B,\qquad \zeta=\frac{S-B}{S+B},\qquad -1\le\zeta\le1,
\end{equation}
be the total distortion and its anisotropic character: $\zeta=1$ where the distortion is a pure splay, $\zeta=-1$ where it is a pure bend,
and $\zeta=0$ where the two are equal. The energy density, as a function of the order at a point where the director field is
given, is then
\begin{equation}\label{eq:frho}
f(\tilde s)=\tilde s^2G-\frac\eta2\,\zeta\,\tilde s^3G+\frac{\lambda^2}8\big(\tilde s^2-1\big)^2 ,
\end{equation}
the three terms being the isotropic elastic energy, the cubic term and the bulk potential; at $\tilde s=1$ it reduces to the
Oseen--Frank density \eqref{eq:OF}. Its derivative is
\begin{equation}
f'(\tilde s)=\Big(2-\frac32\,\eta\,\zeta\,\tilde s\Big)\tilde s\,G+\frac{\lambda^2}2\,\tilde s\,\big(\tilde s^2-1\big),
\end{equation}
and at the bulk value, where the bulk potential has its minimum and contributes nothing,
\begin{equation}\label{eq:frho1}
f'(1)=\Big(2-\frac32\,\eta\,\zeta\Big)G .
\end{equation}
If $f'(1)>0$, a reduction of the order below its bulk value lowers the elastic energy, and the order falls below $1$ until
the Landau bulk term, which grows as $\lambda^2(1-\tilde s)$, stops it: $1-\tilde s\approx(2-\frac32\eta\zeta)G/\lambda^2$. This is the
behaviour of the one-constant model, in which $f'(1)=2G$: distortion always lowers the order. If $f'(1)<0$, an increase of the
order lowers the energy, the cubic term, which grows as $\tilde s^3$, outweighing the isotropic term, which grows as $\tilde s^2$; the
order then rises until the bound stops it, and the bound is active, with the multiplier $\mu$ of \eqref{eq:EL} positive.

The bound is therefore attained only where
\begin{equation}\label{eq:rise}
\eta\,\zeta>\tfrac43 ,
\end{equation}
that is, where the distortion is nearly a pure splay and $\eta>\frac43$, or nearly a pure bend and $\eta<-\frac43$. Since
$|\zeta|\le1$, this is impossible if
\begin{equation}\label{eq:43}
|\eta|<\tfrac43,\qquad\text{that is,}\qquad \tfrac15<\frac{K_3}{K_1}<5 .
\end{equation}
Within this range the bound is not reached outside the cores, and in the cores the order is in any case well below its bulk
value; the equilibria are then expected to be classical solutions of the Euler--Lagrange equations, and the bound serves only to exclude the
fields of \S\ref{sec:negstiff}, of large order and rapid oscillation, which make the unconstrained energy unbounded below. The range
\eqref{eq:43} contains every nematic of Table~\ref{tab:materials}. Three further points follow from \eqref{eq:frho1}. The size
$\lambda$ of the domain enters only through the Landau term, which vanishes at $\tilde s=1$; it sets how far the order falls
below its bulk value, but not the sign of $f'(1)$, so that, at this order, the threshold $\frac43$ does not depend on $\lambda$. The anchoring
does not move it either: with $|\q_b|=\frac12$ and the director tangential, the anchoring energy is least at $\tilde s=1$ and so
contributes nothing to $f'(1)$. Finally, the order is closest to its bulk value where the distortion $G$ is weakest, and is
pushed closer by an anisotropy that favours the local distortion, $\eta\zeta>0$; for $|\eta|<\frac43$ the bound is approached, but
not reached. The criterion \eqref{eq:rise} is local and of leading order: it neglects the gradients of the order, and so
does not amount to a maximum principle for the full equations \eqref{eq:EL}--\eqref{eq:BC}, which would be needed to exclude
the bound in the cores and boundary layers as well. Its predictions are tested in \S\ref{sec:boundres}.

\section{Corners and defects in a large domain}\label{sec:defects}

\subsection{Energies at logarithmic order}

When the domain is large compared with the coherence length, $\lambda\gg1$, and the anchoring is strong, $\gamma\lambda\gg1$, the
order is close to its bulk value except in cores of radius of order $\xi$, that is, of order $1/\lambda$ in our units, at
the defects and the corners; outside them $\q=\frac12(\cos2\varphi,\sin2\varphi)$, the director is tangential on the walls, and
the energy is the Oseen--Frank energy \eqref{eq:OF}. As in the theory of vortices in a superfluid, the energy is then
dominated by terms proportional to $\ln\lambda$, one for each defect and each corner.

These terms are found as follows. Near a defect or a corner, at distances $r$ with $1/\lambda\ll r\ll1$, there is no length
scale, and the director angle depends on the polar angle $\theta$ about the defect or the vertex alone. Then
$\nabla\varphi=r^{-1}\varphi'(\theta)\,\mathbf e_\theta$, and with
\begin{equation}
\psi=\varphi-\theta,
\end{equation}
the angle between the director and the radius,  $\n = \cos\varphi \mathbf e_x+\sin\varphi \mathbf e_y=\cos\psi \mathbf e_r+\sin\psi \mathbf e_\theta$ and $\varphi'=1+\psi'$. The density \eqref{eq:OF} becomes
\begin{align}
&\frac{\varphi'^2}{r^2}\Big[\Big(1-\frac\eta2\Big)\cos^2\psi+\Big(1+\frac\eta2\Big)\sin^2\psi\Big]\nonumber\\
&\qquad=\frac{\varphi'^2}{r^2}\Big[1-\frac\eta2\cos2\psi\Big],
\end{align}
and its integral over $1/\lambda<r<1$ is $\E\ln\lambda$, with
\begin{equation}\label{eq:e}
\E[\psi]=\int\Big(1-\frac\eta2\cos2\psi\Big)(1+\psi')^2\,d\theta ,
\end{equation}
the integral being taken over the wedge of liquid at a corner, or over a full turn at a defect. The coefficient of
$\ln\lambda$ is the least value of $\E$ consistent with the turning of the director. The bracket shows the physics at once:
where the director is radial, $\psi=0$, the distortion is a splay and costs $1-\frac12\eta$; where it is azimuthal,
$\psi=\frac12\pi$, it is a bend and costs $1+\frac12\eta$.

\subsection{The types of a corner}

\begin{figure}[h]
\centering
\includegraphics[width=\columnwidth]{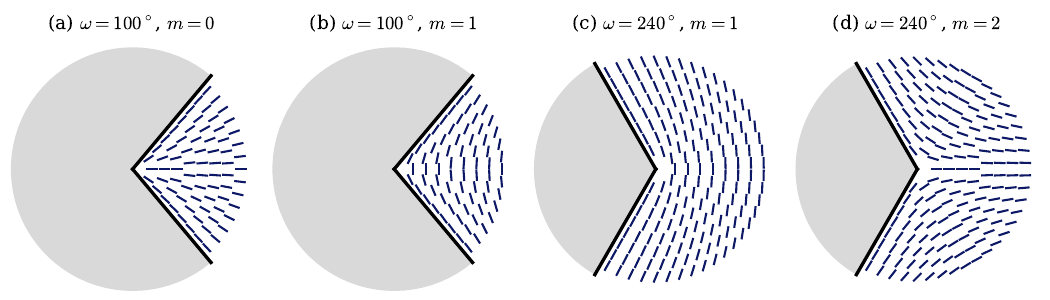}
\caption{Types of a corner of opening angle $\omega$, with tangential anchoring on both walls (thick lines; the solid is
shaded). (a) A splay corner, $m=0$: the director is radial. (b) A half-turn corner, $m=1$: radial on the walls, azimuthal
on the bisector. (c), (d) The reflex corner $\omega=240^\circ$ with $m=1$ and $m=2$. The rods show the profiles
$\psi=-m\pi\theta/\omega$ of least energy in the one-constant model. At $\omega=100^\circ$ and $\omega=240^\circ$ both types shown are
locally stable (Table~\ref{tab:angles}).}
\label{fig:types}
\end{figure}

Consider a corner of opening angle $\omega$, $0<\omega<2\pi$, the liquid occupying the wedge $0<\theta<\omega$ in polar
coordinates about the vertex. On both walls the tangent is radial, so that tangential anchoring requires $\psi=0$ on the
wall $\theta=0$ and $\psi$ equal to a multiple of $\pi$ on the wall $\theta=\omega$, the director being defined modulo $\pi$:
\begin{equation}\label{eq:types}
\psi(0)=0,\qquad \psi(\omega)=-m\pi .
\end{equation}
The integer $m$ is the number of half-turns of the director relative to the radius as the wedge is crossed, and the
director itself turns through $\omega-m\pi$. We call $m$ the \emph{type} of the corner (Fig.~\ref{fig:types}). A corner of
type $m=0$, with no half-turn, is a \emph{splay corner}: the director is radial throughout, fanning out from the vertex like
the spokes of a wheel. A corner of type $m=1$ is a \emph{half-turn corner}: the director is radial on the walls and
azimuthal on the bisector, so that splay and bend alternate. A corner of type $m=2$ is a \emph{full-turn corner}: it has
the least turning only when $\omega>\frac32\pi$, and the director, which turns through two half-turns relative to the radius,
appears to splay towards the vertex. We keep \emph{reflex} for the geometry of a corner, $\omega>\pi$, not for its type. In the one-constant model
the least value of \eqref{eq:e} for type $m$ is attained by the linear profile $\psi=-m\pi\theta/\omega$, and is
\begin{equation}\label{eq:em}
\E_m=\frac{(\omega-m\pi)^2}{\omega}.
\end{equation}
Types $m=0$ and $m=1$ have equal energy at $\omega=90^\circ$, and types $m=1$ and $m=2$ at $\omega=270^\circ$. These are the
angles at which the classification of corners by similarity solutions in the small-domain limit changes from splay to
bend and from bend to its reverse~\cite{RHAM26}.

The types are linked by the defects. Across an arc of radius $r$ about the vertex, the director turns through
\begin{equation}\label{eq:turn}
\Delta\varphi(r)=\varphi(r,\omega)-\varphi(r,0)=\omega+\psi(r,\omega)-\psi(r,0)=\omega-m\pi
\end{equation}
if the corner is of type $m$ at that radius, by \eqref{eq:types} and $\varphi=\theta+\psi$. Now take the closed curve that bounds
the annular sector $r_1<r<r_2$, $0<\theta<\omega$, traversed anticlockwise: out along the wall $\theta=0$ from $r_1$ to $r_2$,
round the arc of radius $r_2$ from $\theta=0$ to $\omega$, back along the wall $\theta=\omega$, and round the arc of radius $r_1$
from $\theta=\omega$ to $0$. Along each straight wall the director is held parallel to the wall and does not turn, so that
the total turning around the curve is $\Delta\varphi(r_2)-\Delta\varphi(r_1)$. This must equal $2\pi$ times the total index $k$ of
the defects inside the sector, the index of a defect being the number of turns of the director, counted anticlockwise, about
it. If the corner is of type $m'$ at $r_1$, close to the vertex, and of type $m$ at $r_2$, far from it, \eqref{eq:turn} gives
$(\omega-m\pi)-(\omega-m'\pi)=2\pi k$, so that the defects between carry total index
\begin{equation}\label{eq:mlink}
k=\tfrac12(m'-m).
\end{equation}
A corner of type $m'$ with a defect of index $+\frac12$ bound to it therefore appears from a distance as a corner of type
$m'-1$, and with a defect of index $-\frac12$ as a corner of type $m'+1$: a half-turn corner is, from afar, a splay corner with
a bound $-\frac12$ defect, and a splay corner a half-turn corner with a bound $+\frac12$ defect.

\subsection{The energies of corners and defects}

\begin{lemma}\label{lem:log}
To leading order in $\ln\lambda$, the energies of corners and defects are $\E\ln\lambda$, with
\begin{enumerate}
\item[(a)] for a splay corner, $m=0$: $\ \E=\omega\big(1-\frac12\eta\big)$, the global least value for $\eta\ge0$ and a strict local
least value for $\eta<0$ and $\omega\le\pi$;
\item[(b)] for a corner of type $m\ge1$: $\ \E\le\E_m$ of \eqref{eq:em}, with equality to first order in $\eta$;
\item[(c)] for a defect of index $\pm\frac12$: $\ \E\le\frac12\pi$, with equality to first order in $\eta$;
\item[(d)] for a defect of index $+1$: $\ \E=2\pi\big(1-\frac12|\eta|\big)$.
\end{enumerate}
\end{lemma}

\begin{proof}
Write $g(\psi)=1-\frac12\eta\cos2\psi$, so that $\E[\psi]=\int g(\psi)\,\varphi'^2\,d\theta$ with $\varphi'=1+\psi'$, and note
$1-\frac12|\eta|\le g\le1+\frac12|\eta|$. Two facts are used repeatedly. Across a corner, $\int_0^\omega\varphi'\,d\theta=\omega-m\pi$ by
\eqref{eq:types}; about a defect of index $k$, $\int_0^{2\pi}\varphi'\,d\theta=2\pi k$. And, by the Cauchy--Schwarz inequality, over an interval of length $L$,
\begin{equation}\label{eq:jensen}
\int\varphi'^2\,d\theta\ \ge\ \frac1L\Big(\int\varphi'\,d\theta\Big)^2 ,
\end{equation}
with equality only when $\varphi'$ is constant.

(a) At a splay corner $\psi=0$ satisfies \eqref{eq:types} and gives $\E=\int_0^\omega g(0)\,d\theta=\omega(1-\frac12\eta)$. If $\eta\ge0$,
then $g\ge1-\frac12\eta$, and \eqref{eq:jensen} with $\int\varphi'\,d\theta=\omega$ gives $\E[\psi]\ge(1-\frac12\eta)\,\omega$ for every
admissible $\psi$: the radial profile is the global minimiser. If $\eta<0$, write $\psi=v$ with $v(0)=v(\omega)=0$ and expand
to second order, using $\cos2v=1-2v^2+O(v^4)$ and $\int v'\,d\theta=0$:
\begin{equation}
\E[v]-\omega\big(1-\tfrac12\eta\big)=\int_0^\omega\Big[\big(1-\tfrac12\eta\big)v'^2+\eta\,v^2\Big]d\theta+O(v^3).
\end{equation}
Since $v$ vanishes at both ends of an interval of length $\omega\le\pi$, $\int v'^2\,d\theta\ge(\pi/\omega)^2\int v^2\,d\theta\ge\int v^2\,d\theta$,
and the bracket is at least $\big[(1+\frac12|\eta|)-|\eta|\big]v^2=(1-\frac12|\eta|)\,v^2>0$ for $|\eta|<2$: the radial profile is
then a strict local minimiser.

(b), (c) In the one-constant model $g=1$, and by \eqref{eq:jensen} the least value of $\E$ is attained by the profile of
constant $\varphi'$, that is, $\psi=a\theta$ with $a=-m\pi/\omega$ across a corner of type $m$ and $a=k-1$ about a defect of index
$k$. Along it $2\psi$ sweeps a whole number of periods, $m$ across the corner and $1$ and $3$ about defects of index
$+\frac12$ and $-\frac12$, so that $\int\cos2\psi\,d\theta=0$ and
\begin{equation}
\E[a\theta]=(1+a)^2\int g(a\theta)\,d\theta=(1+a)^2L
\end{equation}
for every $\eta$, $L$ being $\omega$ or $2\pi$. This equals $\E_m$ at a corner and $\frac12\pi$ at a half-integer defect, and, being the
value of one admissible profile, bounds the least value from above. The least value is stationary with respect to the
profile, so that its derivative with respect to $\eta$ at $\eta=0$ is the explicit derivative of $\E$ on the one-constant
minimiser, \eqref{eq:dEdeta}, which vanishes because $\varphi'$ is constant there and $\int\cos2\psi\,d\theta=0$. The least value
therefore differs from the bound only at order $\eta^2$.

(d) About a defect of index one $\int\varphi'\,d\theta=2\pi$, so that $g\ge1-\frac12|\eta|$ and \eqref{eq:jensen} give
$\E[\psi]\ge2\pi(1-\frac12|\eta|)$. The bound is attained by a constant $\psi$ with $\cos2\psi=\operatorname{sgn}\eta$: a radial director
if $\eta>0$ and an azimuthal one if $\eta<0$. If the walls impose another orientation far from the defect, $\psi$ can change
over an interval of order one in $\ln r$, at a cost independent of $\lambda$, so that the coefficient of $\ln\lambda$ is
unchanged.
\end{proof}

The lemma has a simple interpretation. To first order in $\eta$, the change in the energy of a configuration is its
one-constant profile weighted by the anisotropy of \eqref{eq:e},
\begin{equation}\label{eq:dEdeta}
\frac{d\E}{d\eta}\Big|_{\eta=0}=-\frac12\int\cos2\psi\;\varphi'^2\,d\theta ,
\end{equation}
in which the regions where the director is nearly radial ($\cos2\psi>0$, splay) and nearly azimuthal ($\cos2\psi<0$,
bend) enter with opposite signs. Along the one-constant profiles of a half-integer defect and of a corner of type $m\ge1$,
$\varphi'$ is constant and $2\psi$ sweeps whole periods at a uniform rate, so that the director spends equal parts of the
turn in splay and in bend, the two contributions cancel, and \eqref{eq:dEdeta} vanishes. Only the configurations in which
the director keeps a fixed angle to the radius, the splay corner ($\psi=0$, pure splay) and the defect of index one ($\psi$
constant), have a term linear in $\eta$. The lemma is the first step of an exact solution of the angular problem
\eqref{eq:e}, given in \S\ref{sec:exact}, which shows that the cancellation holds at every odd order; we work with the lemma
first because it gives the critical angles in closed form and shows their structure.

\subsection{How a corner sheds and absorbs a defect}\label{sec:shed}

Consider a corner that appears to be of type $m$ from afar, with a defect of index $k=\pm\frac12$ at a distance $d$ from its
vertex, $1/\lambda\ll d\ll1$. By \eqref{eq:mlink}, the corner is of type $m'=m+2k$ at distances smaller than $d$. In the
annulus $d<r<1$ the field is that of a corner of type $m$; within $r<d$ it is that of a corner of type $m'$ together with
the field of the defect, whose own logarithm runs from its core to a distance of order $d$. The energy is therefore, to
logarithmic accuracy,
\begin{equation}\label{eq:Ed}
E(d)\approx \E_m\ln\frac1d+\Big(\E_{m'}+\frac\pi2\Big)\ln(\lambda d),
\end{equation}
and the defect feels a radial force
\begin{equation}\label{eq:force}
-\frac{dE}{dd}=\frac{\Phi}{d},\qquad \Phi=\E_m-\E_{m'}-\frac\pi2 .
\end{equation}
If $\Phi>0$ the defect is pushed away from the vertex into the liquid: a corner of type $m$ is unstable, and sheds the
defect, leaving a corner of type $m'$. If $\Phi<0$ the defect is drawn in and absorbed, and the corner of type $m$ is
stable against shedding it.

With the energies \eqref{eq:em}, $\Phi=\tfrac 32\pi-(2m+1)\tfrac{\pi^2}{\omega}$ for $k=+\tfrac 12$ and $\Phi=-\tfrac 52\pi+(2m-1)\tfrac{\pi^2}{\omega}$ for $k=-\tfrac 12$. A corner of type $m$ sheds a defect of index $+\frac12$, becoming of type $m+1$, when
$\omega>\frac23(2m+1)\pi$, and sheds a defect of index $-\frac12$, becoming of type $m-1$, when $\omega<\frac25(2m-1)\pi$. A corner of
type $m$ is therefore stable in the window
\begin{equation}\label{eq:window}
\tfrac25(2m-1)\pi<\omega<\tfrac23(2m+1)\pi .
\end{equation}
The resulting critical angles are listed in Table~\ref{tab:angles}. A splay corner is stable up to $120^\circ$ and a half-turn
corner down to $72^\circ$, so that between these angles a corner can be of either type; a half-turn corner remains stable up to
$360^\circ$, and a full-turn corner down to $216^\circ$, so that the reflex corners between $216^\circ$ and $360^\circ$ again
have two stable types. The equal-energy angles $90^\circ$ and $270^\circ$ lie inside these two windows of bistability.

\begin{table}[h]
\small
  \caption{\ Critical angles of a corner in the one-constant model. A corner of type $m$ is stable for $\omega_-<\omega<\omega_+$;
  below $\omega_-$ it sheds a defect of index $-\frac12$, above $\omega_+$ one of index $+\frac12$}
  \label{tab:angles}
  \begin{tabular*}{\columnwidth}{@{\extracolsep{\fill}}lcccc}
    \hline
    Type & Turning & $\E_m$ & $\omega_-$ & $\omega_+$\\
    \hline
    $m=0$, splay & $\omega$ & $\omega$ & -- & $120^\circ$\\
    $m=1$, half-turn &  $\omega-\pi$ & $(\pi-\omega)^2/\omega$ & $72^\circ$ & $360^\circ$\\
    $m=2$, full-turn & $\omega-2\pi$ & $(2\pi-\omega)^2/\omega$ & $216^\circ$ & --\\
    \hline
  \end{tabular*}
\end{table}

Near a critical angle the force \eqref{eq:force} vanishes at leading order, and the position of the defect is set by the
energies of order one, which include its interaction with the walls. A defect may then be held at a finite distance from
the vertex. This is the \emph{hidden} defect of Refs.~\cite{Yao22,RHAM26}: in an isosceles triangle with Dirichlet
anchoring, the apex changes from a splay profile with a nearby $-\frac12$ defect to a half-turn profile at an apex angle of about
$72^\circ$ in the Oseen--Frank limit, and of about $74^\circ$ at $\lambda=100$~\cite{RHAM26}. That is the angle $\omega_-=72^\circ$ at
which a half-turn corner sheds a $-\frac12$ defect. The same angle bounds the trefoil state found numerically in those
triangles~\cite{RHAM26}, which has three splay corners and an interior $-\frac12$ defect: at $\lambda=100$ it is the minimiser only
while every vertex angle lies below about $72^\circ$, giving way to a rotated state with a bend apex when the apex angle exceeds
$74^\circ$, and to rotated states with a bend base vertex when the base angles exceed $70.5^\circ$ (apex angles below $39^\circ$). Both
thresholds lie within $2^\circ$ of $\omega_-$, the small differences being due to the energies of order one and to the Dirichlet
anchoring used there; \S\ref{sec:irregular} derives both thresholds from the energies of the corners. Its counterpart at $\omega_+=120^\circ$, a $+\frac12$ defect drawn into or released
from a splay corner, governs the polygons of more than five sides, and it is this angle that the anisotropy moves.

\paragraph{The effect of the anisotropy.} By lemma~\ref{lem:log}, to first order in $\eta$ only the energy of the splay
corner changes, to $\E_0=\omega(1-\frac12\eta)$; those of the half-turn corner, $\E_1=(\pi-\omega)^2/\omega$, and of the half-integer
defects, $\frac12\pi$, are unchanged. A critical angle is where the force \eqref{eq:force} on a defect at the vertex vanishes,
$\Phi=0$. For a half-turn corner shedding a $-\frac12$ defect and for a splay corner shedding a $+\frac12$ defect these
conditions are $\E_1=\E_0+\frac12\pi$ and $\E_0=\E_1+\frac12\pi$, that is,
\begin{equation}\label{eq:quadratics}
\begin{split}
&\frac{(\pi-\omega)^2}{\omega}=\omega\Big(1-\frac\eta2\Big)\pm\frac\pi2\\
&\quad\Longleftrightarrow\quad
\eta\,\omega^2-(4\pm1)\pi\,\omega+2\pi^2=0 ,
\end{split}
\end{equation}
with the upper sign for the first and the lower for the second. Of the two roots of each quadratic we take the one that
tends to the one-constant value as $\eta\to0$: a half-turn corner sheds a $-\frac12$ defect below
\begin{equation}\label{eq:omm}
\omega_-(\eta)=\frac\pi\eta\Big[\frac52-\Big(\frac{25}4-2\eta\Big)^{1/2}\Big]\ \to\ \frac{2\pi}5 ,
\end{equation}
and a splay corner sheds a $+\frac12$ defect above
\begin{equation}\label{eq:omp}
\omega_+(\eta)=\frac\pi\eta\Big[\frac32-\Big(\frac94-2\eta\Big)^{1/2}\Big]\ \to\ \frac{2\pi}3 .
\end{equation}
The two types have equal energy where $\E_0=\E_1$, that is, $\omega(1-\frac12\eta)^{1/2}=\pi-\omega$, or
$\omega_=(\eta)=\pi/[1+(1-\frac12\eta)^{1/2}]\to\frac12\pi$. The critical angles of reflex corners involve only half-turn and
full-turn corners, whose energies are unchanged at this order, and so do not move. Table~\ref{tab:anis} gives the three
angles over the range of the computations, both from these formulae and exactly (\S\ref{sec:exact}). All three increase with
$\eta$: when bend is the more expensive distortion, splay corners survive to wider angles. The angle $\omega_+$ is the most
sensitive; exactly, it moves from $104^\circ$ to $136^\circ$ as $K_3/K_1$ increases from $0.54$ to $1.86$, and the first-order
formula \eqref{eq:omp} overestimates the movement, by $3^\circ$ at $\eta=-0.6$ and $7^\circ$ at $\eta=0.6$.

\begin{table}[h]
\footnotesize\setlength{\tabcolsep}{2pt}
  \caption{\ Critical angles of a corner, in degrees, for the anisotropies of the computations: to first order in $\eta$,
  from \eqref{eq:omm}, \eqref{eq:omp} and $\omega_=(\eta)$, and exactly (\S\ref{sec:exact})}
  \label{tab:anis}
  \begin{tabular*}{\columnwidth}{@{\extracolsep{\fill}}llcccccc}
    \hline
    & $\eta$ & $-0.6$ & $-0.2$ & $0$ & $0.2$ & $0.4$ & $0.6$\\
    & $K_3/K_1$ & $0.54$ & $0.82$ & $1$ & $1.22$ & $1.50$ & $1.86$\\
    \hline
    $\omega_-$ & 1st order & $68.8$ & $70.9$ & $72.0$ & $73.2$ & $74.5$ & $75.8$\\
     & exact & $68.7$ & $70.9$ & $72.0$ & $73.2$ & $74.4$ & $75.7$\\
    $\omega_=$ & 1st order & $84.1$ & $87.9$ & $90.0$ & $92.4$ & $95.0$ & $98.0$\\
     & exact & $83.5$ & $87.8$ & $90.0$ & $92.3$ & $94.8$ & $97.4$\\
    $\omega_+$ & 1st order & $107.2$ & $115.1$ & $120.0$ & $125.9$ & $133.1$ & $142.6$\\
     & exact & $103.9$ & $114.6$ & $120.0$ & $125.3$ & $130.5$ & $135.7$\\
    \hline
  \end{tabular*}
\end{table}

\subsection{Polygons and the disc}\label{sec:polygons}

The topology of the boundary fixes the total index of the defects in the interior. The following is the defect rule of Yao
\emph{et al.}~\cite{Yao22}, stated for the corner types of \S\ref{sec:shed}; it holds for any simple polygon, convex or not.

\begin{proposition}\label{prop:index}
With strong tangential anchoring, the indices $k_i$ of the defects in a polygon whose corners are of types $m=0$, $1$ or $2$
sum to
\begin{equation}\label{eq:indexgen}
\sum_ik_i=1-\tfrac12\big(n_0-n_2\big),
\end{equation}
$n_0$ being the number of splay corners and $n_2$ the number of full-turn corners. In the disc they sum to one.
\end{proposition}

\begin{proof}
Remove a small disc about each defect. The director angle changes by zero around the boundary of what remains, for
$\nabla\varphi$ is defined there and has zero circulation. Traverse the outer boundary anticlockwise. Along the edges the director
does not turn. At a corner of interior angle $\omega_i$ the tangent turns through $\pi-\omega_i$, and by \eqref{eq:turn} the director turns
through $\pi-\omega_i-(1-m_i)\pi$: with the tangent at a half-turn corner, by $\pi$ less at a splay corner, and by $\pi$ more at a
full-turn corner. Since the tangent turns through $2\pi$ in all, the total is $2\pi-(n_0-n_2)\pi$. Around the defects, traversed in
the opposite sense, the director turns through $-2\pi\sum k_i$. In the disc the director follows the tangent, which turns through
$2\pi$.
\end{proof}

In a convex polygon every corner is a splay or a half-turn corner (Table~\ref{tab:angles}), $n_2=0$, and \eqref{eq:indexgen} becomes
\begin{equation}\label{eq:index}
\sum_ik_i=1-\tfrac12n_s ,
\end{equation}
with $n_s=n_0$ the number of splay corners. The interior angle of a regular $N$-gon is $\omega_N=\pi-2\pi/N$, so that this is the
case of the regular polygons.

The proposition assumes that the director is tangential up to the corners. Within a distance of order $l_{\rm ex}$ of a corner
it is not, and a corner can hold a surface defect that partly undoes the turning~\cite{Raj26a}; at $\gamma\lambda=100$ this distance
is a hundredth of the size of the domain. A defect bound to a corner, in the sense of \S\ref{sec:shed}, changes the type
of the corner and so the count $n_s$; in \eqref{eq:index} it is counted with the corner, not with the interior defects.
The square with $n_s=2$, the splay corners at opposite ends of a diagonal, is the \emph{diagonal} state, which has no
defects; with $n_s=0$ the square must hold defects of total index $+1$, and with $n_s=4$ of total index $-1$. A polygon thus
admits, at leading order, states with $n_s=0,1,\dots,N$, and for each $n_s$ several arrangements of the splay corners, which
differ in energy only at order one and are a source of multistability.

At leading order in $\ln\lambda$ the energy of a state is the sum of the energies of its corners
and of its interior defects, each of the form (coefficient)$\times\ln\lambda$; their interactions
contribute only at order one. A state with $n_s$ splay corners contains three kinds of element:
\begin{enumerate}
\item $n_s$ splay corners, each with coefficient $\omega_N(1-\frac12\eta)$ (lemma~\ref{lem:log});
\item $N-n_s$ half-turn corners, each with coefficient $(\pi-\omega_N)^2/\omega_N=4\pi/[N(N-2)]$,
which is unaffected by the anisotropy at first order in $\eta$;
\item $2|1-\frac12n_s|$ interior defects of index $\frac12\operatorname{sgn}(1-\frac12n_s)$, each with
coefficient $\frac12\pi$, likewise unaffected at first order.
\end{enumerate}
The number of defects in item 3 follows from \eqref{eq:index}. The defects must carry the total
index $1-\frac12n_s$, and since a defect of index $k$ costs $2\pi k^2$ when $\eta=0$, an index $k$
is carried most cheaply by $2|k|$ defects of index $\frac12\operatorname{sgn}k$, which cost
$\pi|k|\le2\pi k^2$; a further pair of defects of opposite sign would add $\pi$ without changing
the index. (For the total index $+1$, that is $n_s=0$, splitting is favourable at first order
when $|\eta|<1$, as for the disc below.) Adding the three contributions,
\begin{equation*}
\begin{split}
E=\Big[\,&n_s\,\omega_N\Big(1-\frac\eta2\Big)+(N-n_s)\frac{(\pi-\omega_N)^2}{\omega_N}\\
&+2\Big|1-\frac{n_s}2\Big|\cdot\frac\pi2\,\Big]\ln\lambda ,
\end{split}
\end{equation*}
and writing $E=\pi c_{n_s}\ln\lambda$ with $\omega_N=\pi(1-2/N)$ gives
\begin{equation}\label{eq:cn}
\begin{split}
c_{n_s}={}&\underbrace{n_s\Big(1-\frac2N\Big)\Big(1-\frac\eta2\Big)}_{\text{splay corners}}
+\underbrace{\frac{4(N-n_s)}{N(N-2)}}_{\text{half-turn corners}}\\
&+\underbrace{\Big|1-\frac {n_s}2\Big|}_{\text{interior defects}} .
\end{split}
\end{equation} Table~\ref{tab:cn} gives the values for
$\eta=0$, and three results follow.

\begin{table*}[!tb]
\small
  \caption{\ Coefficients $c_{n_s}$ of $\pi\ln\lambda$ in the energy of a state with $n_s$ splay corners, for $\eta=0$, with
  the first-order corrections $-\frac12n_s(1-2/N)\eta$ in brackets; the least value in each row is in bold, and $\eta_c$ is the
  anisotropy at which the states with $n_s=0$, $1$ and $2$ exchange, to first order \eqref{eq:etac} and exactly
  (\S\ref{sec:exact})}
  \label{tab:cn}
  \begin{tabular*}{\textwidth}{@{\extracolsep{\fill}}ccccccccc}
    \hline
    $N$ & $\omega_N$ & $n_s=0$ & $n_s=1$ & $n_s=2$ & $n_s=3$ & $n_s=4$ & $\eta_c$ & $\eta_c$, exact\\
    \hline
    3 & $60^\circ$ & $5$ & $3.50\ (-0.17\eta)$ & $2\ (-0.33\eta)$ & $\mathbf{1.50}\ (-0.50\eta)$ & -- & -- & -- \\
    4 & $90^\circ$ & $3$ & $2.50\ (-0.25\eta)$ & $\mathbf{2}\ (-0.50\eta)$ & $2.50\ (-0.75\eta)$ & $3\ (-\eta)$ & $-2$ & -- \\
    5 & $108^\circ$ & $2.33$ & $2.17\ (-0.30\eta)$ & $\mathbf{2}\ (-0.60\eta)$ & $2.83\ (-0.90\eta)$ & $3.67\ (-1.20\eta)$ & $-0.56$ & $-0.45$ \\
    6 & $120^\circ$ & $\mathbf2$ & $\mathbf2\ (-0.33\eta)$ & $\mathbf2\ (-0.67\eta)$ & $3\ (-\eta)$ & $4\ (-1.33\eta)$ & $0$ & $0$ \\
    7 & $128.6^\circ$ & $\mathbf{1.80}$ & $1.90\ (-0.36\eta)$ & $2\ (-0.71\eta)$ & $3.10\ (-1.07\eta)$ & $4.20\ (-1.43\eta)$ & $0.28$ & $0.32$ \\
    8 & $135^\circ$ & $\mathbf{1.67}$ & $1.83\ (-0.38\eta)$ & $2\ (-0.75\eta)$ & $3.17\ (-1.13\eta)$ & $4.33\ (-1.50\eta)$ & $0.44$ & $0.57$ \\
    \hline
  \end{tabular*}
\end{table*}

First, the state with two splay corners and no defects has $c_2=2-(1-2/N)\eta$, and in particular the energy $2\pi\ln\lambda$
when $\eta=0$, \emph{whatever the number of sides}. The two splay corners turn the director through $2\omega_N$ and the half-turn
corners through the rest, and their contributions, $2\omega_N=2\pi-4\pi/N$ and $(N-2)(\pi-\omega_N)^2/\omega_N=4\pi/N$, sum to $2\pi$.

\begin{figure}[h]
\centering
\includegraphics[width=\columnwidth]{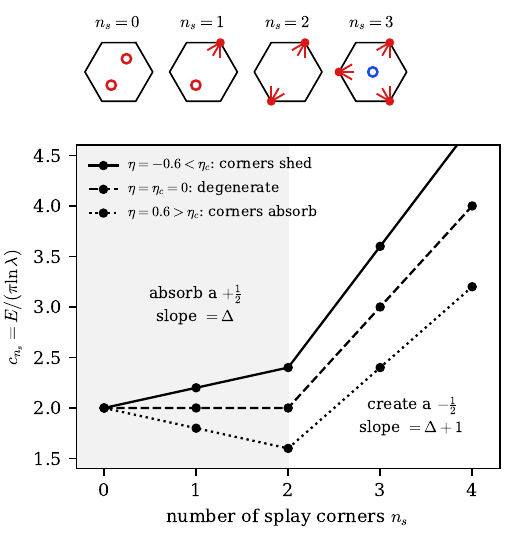}
\caption{The energy of a polygon as its corners are converted one at a time from half-turn to splay corners, for the hexagon
($N=6$, $\eta_c=0$), from \eqref{eq:cn}. Top: the states, with splay corners marked by the fanning director, interior defects of
index $+\frac12$ red and of index $-\frac12$ blue. Up to $n_s=2$ each conversion absorbs a $+\frac12$ defect, and the energy changes by
the same step $\Delta$ \eqref{eq:step}; beyond $n_s=2$ each conversion must create a $-\frac12$ defect, and the step is $\Delta+1>0$.
The least energy is therefore at $n_s=0$ when $\Delta>0$ ($\eta<\eta_c$, solid) and at $n_s=2$ when $\Delta<0$ ($\eta>\eta_c$, dotted);
at $\eta=\eta_c$ (dashed) the states with $n_s=0$, $1$ and $2$ are degenerate.}
\label{fig:ladder}
\end{figure}

Secondly, which state does a polygon prefer? The answer is clearest if we follow the energy as the corners are
converted one at a time from half-turn to splay corners (Fig.~\ref{fig:ladder}). By \eqref{eq:index}, each conversion lowers
the total index of the interior defects by $\frac12$. While that index is positive, $n_s\le1$, the interior holds $+\frac12$
defects, and the conversion removes one of them: the corner absorbs it. Read in reverse, this is the shedding of
\S\ref{sec:shed}, in which a splay corner emits a $+\frac12$ defect into the liquid and becomes a half-turn corner. The change
in energy is the cost of one splay corner, less the costs of the half-turn corner it replaces and of the free $+\frac12$
defect it absorbs. In units of $\pi\ln\lambda$ this step is
\begin{equation}\label{eq:step}
\begin{split}
\Delta&=c_{n_s+1}-c_{n_s}\\
&=\frac1\pi\Big[\omega_N\Big(1-\frac\eta2\Big)-\frac{(\pi-\omega_N)^2}{\omega_N}-\frac\pi2\Big]\quad(n_s\le1).
\end{split}
\end{equation}
It does not depend on $n_s$, for at leading order each corner contributes independently, and converting a second corner
costs exactly as much as converting the first. The energy is therefore a straight line in $n_s$ from $n_s=0$ to $n_s=2$, of
slope $\Delta$, and $c_1=\frac12(c_0+c_2)$ lies midway between its neighbours.

At $n_s=2$ the interior holds no defects, and the rule changes. A further conversion can no longer absorb a $+\frac12$ defect,
for none is left; to lower the interior index by $\frac12$ it must instead create a defect of index $-\frac12$, which costs
$\frac12\pi\ln\lambda$ rather than saving it. The step beyond $n_s=2$ is therefore $\Delta+1$, and this is positive for every polygon
with $N\ge4$ and every $|\eta|<2$. The energy thus falls or rises linearly up to $n_s=2$ and then rises steeply
(Fig.~\ref{fig:ladder}), so that its least value is attained at one end of the straight segment, $n_s=0$ or $n_s=2$, and the
sign of $\Delta$ decides which. If $\Delta>0$, a splay corner costs more than the half-turn corner and the free defect it
replaces; every corner sheds its defect, and the preferred state has $n_s=0$ and two $+\frac12$ defects in the interior. If
$\Delta<0$, the corners absorb the defects, and the preferred state has $n_s=2$ and none. The state with one splay corner lies
on the segment between them and is never preferred at this order; when $\Delta=0$ the three states are degenerate, and the
energies of order one decide among them. The triangle is the exception: there $\Delta=-\frac32-\frac16\eta$ and $\Delta+1=-\frac12-\frac16\eta<0$ for every $|\eta|<2$, so
that even a conversion that creates a $-\frac12$ defect lowers the energy, and the energy falls all the way to $n_s=3$: the
triangle prefers three splay corners and a $-\frac12$ defect, as its angle, below $\omega_-$, requires.

The step vanishes when
\[
\omega_N\Big(1-\frac\eta2\Big)=\frac{(\pi-\omega_N)^2}{\omega_N}+\frac\pi2 .
\]
This is the condition that defined $\omega_+(\eta)$ in \S\ref{sec:shed}, where a single corner of angle $\omega$ was found
to be indifferent between keeping its splay and shedding a $+\frac12$ defect. Because the energy is a sum over the
corners, nothing global is involved: the polygon exchanges its states exactly when each of its corners reaches this local
threshold, $\omega_N=\omega_+(\eta)$. To find the anisotropy at which this happens for a regular $N$-gon, it is simplest
to equate the two competing states, $c_0=(N+2)/(N-2)$ and $c_2=2-(1-2/N)\eta$, which is equivalent by linearity:
\[
\Big(1-\frac2N\Big)\eta=2-\frac{N+2}{N-2}=\frac{N-6}{N-2},
\]
whence the critical anisotropy
\begin{equation}\label{eq:etac}
\eta_c=\frac{N(N-6)}{(N-2)^2}.
\end{equation}
Since $c_2$ decreases with $\eta$ and $c_0$ does not depend on it, the polygon prefers two splay corners and no defects for
$\eta>\eta_c$, and two $+\frac12$ defects in the interior for $\eta<\eta_c$. The critical anisotropy vanishes for the hexagon, whose angle is exactly $120^\circ$, so that there the sign of
$K_3-K_1$ decides the state; it is $-0.56$ for the pentagon, within reach of bent-core nematics, and $0.28$ and $0.44$ for
the heptagon and octagon, within reach of 5CB. The exact values (\S\ref{sec:exact}) are $-0.45$, $0$, $0.32$ and $0.57$.

Thirdly, the triangle and the square are far from $\omega_+$. The angle of the triangle, $60^\circ$, is below $\omega_-$ for every
$\eta$ in Table~\ref{tab:anis}: no half-turn corner can survive, all three corners are splay, and by \eqref{eq:index} the
interior holds a defect of index $-\frac12$, which the symmetry places at the centre. The angle of the square, $90^\circ$, is
within the window of bistability, both types of corner are possible, and \eqref{eq:cn} is least for $n_s=2$ whenever
$|\eta|<2$.

\paragraph{The disc.} In the disc the defects carry total index one, either as a single defect of index $+1$ or as two of
index $+\frac12$. By lemma~\ref{lem:log} their coefficients are $2\pi(1-\frac12|\eta|)$ and at most $\pi$, so that the split state
has the lower energy at leading order whenever $|\eta|<1$, that is, whenever $\frac13<K_3/K_1<3$; the exact energies of
\S\ref{sec:exact} extend this to $|\eta|\le1.95$, practically the whole admissible range. The defect of index one, whose director can adopt everywhere the cheaper of the two pure distortions, the bend of a
circular field or the splay of a radial one, gains from the anisotropy at first order; the two half-integer defects gain
only at second order, but gain more. As $N\to\infty$, $\eta_c\to1$: the two
splay corners of a polygon with many sides become the two halves of a $+1$ defect pushed to the wall.

\subsection{Irregular polygons}\label{sec:irregular}

Nothing in the argument of \S\ref{sec:polygons} requires the polygon to be regular. In a convex polygon with interior angles
$\omega_1,\dots,\omega_N$, each corner is a splay or a half-turn corner, \eqref{eq:index} still fixes the interior index at
$1-\frac12n_s$, and the energy at leading order is still a sum over the corners and the interior defects:
\begin{equation}\label{eq:Eirr}
\begin{split}
\frac{E}{\ln\lambda}={}&\sum_{i=1}^N\Big[\sigma_i\,\E_0(\omega_i)+(1-\sigma_i)\,\E_1(\omega_i)\Big]\\
&+n_+\E_{+1/2}+n_-\E_{-1/2} ,
\end{split}
\end{equation}
where $\sigma_i=1$ if corner $i$ is a splay corner and $0$ if it is a half-turn corner, $n_s=\sum_i\sigma_i$, and the interior holds
$n_+=\max(2-n_s,0)$ defects of index $+\frac12$ and $n_-=\max(n_s-2,0)$ of index $-\frac12$. To first order in $\eta$,
$\E_0(\omega)=\omega(1-\frac12\eta)$, $\E_1(\omega)=(\pi-\omega)^2/\omega$ and $\E_{\pm1/2}=\frac12\pi$, so that the defect term is
$\pi|1-\frac12n_s|$ as in \eqref{eq:cn}; with the exact energies of \S\ref{sec:exact} for all four, the argument below is unchanged and
gives the exact critical angles. The energy now depends on which
corners are splay, and not only on how many, but the minimisation is no harder. Converting corner $i$ from half-turn to
splay changes the energy of the corners by $D(\omega_i)=\E_0(\omega_i)-\E_1(\omega_i)$, which increases with $\omega$ for $|\eta|<2$: sharp
corners gain most from being splay. The defect term changes by $-\E_{+1/2}$ for each of the first two conversions, which absorb
$+\frac12$ defects, and by $+\E_{-1/2}$ for each later one, which creates a $-\frac12$ defect (Fig.~\ref{fig:ladder}). The least energy is
therefore found by converting the corners in order of increasing angle for as long as each conversion lowers the energy. A
conversion that absorbs a $+\frac12$ defect does so when $D(\omega)<\E_{+1/2}$, and one that creates a $-\frac12$ defect when
$D(\omega)<-\E_{-1/2}$; these are precisely the conditions $\omega<\omega_+$ and $\omega<\omega_-$ of \S\ref{sec:shed}. The selection rule is thus:
\emph{the two sharpest corners are splay corners if their angles lie below $\omega_+$; every further corner is a splay corner
only if its angle lies below $\omega_-$.} For a regular polygon it reproduces \S\ref{sec:polygons}. The criterion is local, for the energy of each
corner depends on its own angle alone; but the corners are coupled through the topological constraint \eqref{eq:index},
which fixes how many defects the interior must hold, and so the cost of a conversion depends on how many corners have
already been converted. This is why a corner's rank among the angles, and not only its angle, decides its type.

For a triangle, whose angles sum to $\pi$, the two sharpest angles are always below $90^\circ<\omega_+$, so that two corners at
least are splay, and the rule reduces to a single test: the third, widest, corner is a splay corner, and the interior holds a
$-\frac12$ defect, if and only if its angle lies below $\omega_-$. A triangle all of whose angles lie below $\omega_-$ takes the
trefoil state; otherwise its widest corner is a half-turn corner and the interior holds no defect. In an isosceles triangle of
apex angle $\alpha$ and base angles $\frac12(\pi-\alpha)$, the widest corner is the apex when $\alpha>60^\circ$ and a base vertex when
$\alpha<60^\circ$, so that the trefoil state is preferred for
\begin{equation}\label{eq:isosceles}
\pi-2\omega_-<\alpha<\omega_- ,
\end{equation}
that is, $36^\circ<\alpha<72^\circ$ in the one-constant model; for $\alpha>\omega_-$ the apex, and for $\alpha<\pi-2\omega_-$ one of the base
vertices, is a half-turn corner. At $\lambda=100$ the computations of Ref.~\cite{RHAM26} place these transitions at $74^\circ$ and
$39^\circ$, the second corresponding to base angles of $70.5^\circ$, each within $3^\circ$ of the prediction. The anisotropy widens the
window when $\eta>0$ and narrows it when $\eta<0$: with the exact values of $\omega_-$ in Table~\ref{tab:anis}, \eqref{eq:isosceles} gives
$42.6^\circ<\alpha<68.7^\circ$ at $\eta=-0.6$ and $28.6^\circ<\alpha<75.7^\circ$ at $\eta=0.6$.

A polygon with re-entrant corners is treated in the same way, with \eqref{eq:indexgen} in place of \eqref{eq:index}. A
reflex corner, $\pi<\omega<2\pi$, is a half-turn or a full-turn corner, and converting it from the first to the second raises the
interior index by $\frac12$: it absorbs a $-\frac12$ defect if one is present, and must otherwise create a $+\frac12$ defect. The
same argument suggests that the conversion lowers the energy only if a $-\frac12$ defect is available to be absorbed and the
angle exceeds $216^\circ$ (Table~\ref{tab:angles}); creating a $+\frac12$ defect would require an angle above $360^\circ$, which a corner
cannot have. Whether a given re-entrant corner is in fact a full-turn corner depends on the charge left in the interior by
the other corners, and on the competing arrangements, so that we state this as a prediction rather than a rule: re-entrant
corners wider than $216^\circ$ should act as sinks for the $-\frac12$ defects that sharp corners shed, which star-shaped or L-shaped
wells could test.

The rule selects the type of each corner, and so the number and sign of the interior defects, at leading order. It cannot
distinguish arrangements of the splay corners among corners of equal angle, nor resolve near-degeneracies close to $\omega_\pm$;
both are decided by the energies of order one, as in the regular polygons.

These conclusions hold at leading order in $\ln\lambda$. At $\lambda=100$, $\ln\lambda\approx4.6$, and the energies of order one, which
depend on the positions of the defects and on the arrangement of the splay corners, are not negligible. The corrections
of higher order in $\eta$, which are not negligible either when $|\eta|$ is about one half, are found next.

\subsection{Beyond first order in the anisotropy}\label{sec:exact}

The results so far rest on the first-order energies of lemma~\ref{lem:log}. The angular problem \eqref{eq:e} can however be
solved exactly, and the solution shows both why the first-order results have the structure they do and how large the
corrections are. Write
\begin{equation}
g(\psi)=1-\tfrac12\eta\cos2\psi
\end{equation}
for the stiffness of a distortion in which the director makes the angle $\psi$ with the radius. The integrand
$g(\psi)(1+\psi')^2$ of \eqref{eq:e} does not depend on $\theta$ explicitly, and so, just as the energy of a pendulum is
conserved because its Lagrangian does not depend on time, its Euler--Lagrange equation has a first integral,
\begin{equation}\label{eq:firstint}
g(\psi)\,(\psi'^2-1)=C .
\end{equation}
Across a corner of type $m\ge1$, or about a defect of index $\pm\frac12$, $\psi$ decreases through $n$ half-turns, with $n=m$
at a corner and $n=1$ and $3$ about defects of index $+\frac12$ and $-\frac12$, over the angle $L$ available, $L=\omega$ at a
corner and $2\pi$ about a defect. Then $\psi'=-\sigma(\psi)$, with $\sigma=(1+C/g)^{1/2}$, so that $d\theta=d\psi/\sigma$ and
$1+\psi'=1-\sigma$. The constant $C$ is fixed by requiring that the $n$ half-turns fill the angle $L$, and the coefficient of
$\ln\lambda$ then follows by quadrature:
\begin{equation}\label{eq:quad}
n\int_0^\pi\frac{d\psi}{\sigma}=L,\qquad \E=n\int_0^\pi g\,\frac{(1-\sigma)^2}{\sigma}\,d\psi .
\end{equation}
We have checked these values against a direct numerical minimisation of \eqref{eq:e}.

Two consequences are immediate. First, because $\psi$ sweeps whole half-turns, the integrals \eqref{eq:quad} are unchanged by
the shift $\psi\mapsto\psi+\frac12\pi$, which changes $\eta$ into $-\eta$: the energies of the corners of type $m\ge1$ and of the
half-integer defects are \emph{even} functions of $\eta$. Their first-order corrections in lemma~\ref{lem:log} therefore vanish
exactly, and the anisotropy enters them first at second order. The splay corner and the defect of index one, in which $\psi$
is constant, are the only configurations whose energy has a term linear in $\eta$. Secondly, expanding \eqref{eq:quad} for
small $\eta$, or equivalently perturbing the linear profile $\psi=a\theta$, gives closed forms with errors of order $\eta^4$,
\begin{gather}
\E_m=\frac{(\omega-m\pi)^2}{\omega}-\eta^2\,\frac{(\omega^2-m^2\pi^2)^2}{32\,m^2\pi^2\,\omega},\label{eq:em2}\\
\E_{+1/2}=\frac\pi2\Big(1-\frac{9}{32}\eta^2\Big),\qquad \E_{-1/2}=\frac\pi2\Big(1-\frac{25}{288}\eta^2\Big).\label{eq:eh2}
\end{gather}
At $|\eta|=0.6$ these differ from the exact values by less than $0.2$ per cent. The correction to a half-turn corner vanishes as
$\omega\to\pi$, where the corner disappears, and is small at the angles of the polygons. A defect of index $+\frac12$ relaxes
more than three times as much as one of index $-\frac12$: about the first the director turns slowly relative to the radius,
$\psi'=-\frac12$, and can linger near the cheaper orientation; about the second it turns three times as fast.

The corrections leave the structure of \S\S\ref{sec:shed}--\ref{sec:polygons} intact, and alter its numbers in one place. The
angles $\omega_-$ and $\omega_=$ move by less than a degree (Table~\ref{tab:anis}), for they involve the $-\frac12$ defect and the
half-turn corner, whose corrections are small. The angle $\omega_+$ involves the $+\frac12$ defect, which becomes cheaper for either
sign of $\eta$, so that a splay corner sheds it more readily on both sides: exactly, $\omega_+$ ranges from $103.9^\circ$ to
$135.7^\circ$ over the computations, against $107.2^\circ$ to $142.6^\circ$ at first order. The critical anisotropies become
$-0.45$, $0$, $0.32$ and $0.57$ for the pentagon to the octagon (Table~\ref{tab:cn}); that of the hexagon remains zero at
every order, because the corrections vanish at $\eta=0$. The second-order forms \eqref{eq:em2}--\eqref{eq:eh2} reproduce
these values to within $0.003$. In the disc, twice $\E_{+1/2}$ lies below $2\pi(1-\frac12|\eta|)$ for every $|\eta|\le1.95$ (both tend to zero as
$|\eta|\to2$), so that at leading order in $\ln\lambda$ the defect of index one is never preferred.

The first-order truncation is therefore more than a convenience: it gives the critical angles in closed form, it identifies
correctly the one configuration, the splay corner, through which the anisotropy acts at first order, and it is exact for
that configuration. The higher orders change the numbers by a few degrees, and only where a $+\frac12$ defect is involved.
All of these results remain, of course, of leading order in $\ln\lambda$.

\section{Numerical method}\label{sec:numerics}

The energy \eqref{eq:F} is discretised with continuous piecewise-linear finite elements on an unstructured triangulation,
with mesh size $h=0.72/\lambda$, about three elements per coherence length, so that the meshes have between $3\times10^4$
and $7\times10^4$ nodes. On a triangle the gradient is constant and $\q$ is linear, so that the integrand of the cubic term
is linear and its integral is the area times its value at the centroid: the cubic term is integrated exactly, and its
gradient and Hessian are assembled analytically (Appendix~\ref{app:discrete}). The Landau energy is integrated with a
lumped mass matrix and the anchoring energy with the trapezoidal rule on each boundary edge, using the tangent of that
edge, so that the corners need no special treatment.

The bound \eqref{eq:bound} is imposed at the nodes. Stationary points are found by a Newton iteration with an active set
(Appendix~\ref{app:newton}): nodes at which the bound is attained and the energy would decrease by increasing the order are
held on the bound, and the iterate is projected onto the bound by \eqref{eq:proj} after each step. Each step is damped so
that the energy decreases (a trust-region, or Levenberg--Marquardt, iteration), and the iteration converges when the
residual of the first-order conditions is below $10^{-6}$. An equilibrium is accepted as stable if the Hessian of the
energy, restricted to the perturbations that respect the active constraints, is positive definite, which is tested
exactly by a Cholesky factorisation; saddle points are displaced along their softest mode and minimised again.

The equilibria are sought from many initial fields: a uniform director at eight orientations; a director parallel to the
nearest edge; circular and radial fields; fields with prescribed defects of index $\pm\frac12$ at random positions and of total
index between $-1$ and $1$; and random smooth fields. They are first relaxed on a coarser mesh, $h=1.5/\lambda$, and the
distinct results refined on the fine mesh. Two states are identified if they have the same number of interior defects
and the same total index, energies within $5\times10^{-4}$ of each other relative to the energy, the same macroscopic order,
and average order tensors related by a symmetry of the domain. Every stable state is continued in $\eta$ over the interval
$[-0.6,0.6]$, in steps of $0.2$, until it loses stability or merges with another, and the search is repeated at every value
of $\eta$.

The defects are located as the connected regions in which $s/s_+<\frac12$, and the index of each is computed from the
winding of $2\varphi$ around the triangles of the mesh. A defect whose centre lies within $5\xi$ of the wall is regarded as
bound to it, and so, by \S\ref{sec:shed}, as part of a corner; the number of splay corners then follows from
\eqref{eq:index} applied to the interior defects alone. The macroscopic order of a state is
\begin{equation}
\bar s=2|\langle\q\rangle|,
\end{equation}
where $\langle\q\rangle$ is the average of $\q$ over the domain; $\bar s=1$ for a uniform nematic of bulk order, and $\bar s=0$
for a state whose average order tensor vanishes. 

\begin{figure}[h]
\centering
\includegraphics[width=\columnwidth]{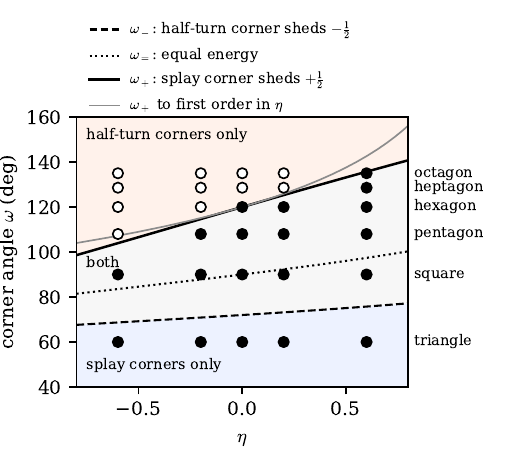}
\caption{Critical angles of a corner against the anisotropy $\eta$: below $\omega_-$ (dashed) a half-turn corner sheds a defect of
index $-\frac12$, so that only splay corners survive; above $\omega_+$ (solid) a splay corner sheds a defect of index
$+\frac12$, so that only half-turn corners survive; between them both types are stable, and they have equal energy on $\omega_=$
(dotted). These curves are exact at leading order in $\ln\lambda$ (\S\ref{sec:exact}); the grey curve is $\omega_+$ to first
order in $\eta$, \eqref{eq:omp}. The horizontal lines mark the interior angles of the polygons. The symbols are the computed global minimisers at
$\lambda=100$: filled, the minimiser has splay corners ($n_s=2$, or $n_s=3$ in the triangle); open, it has none, and holds two
interior defects of index $+\frac12$.}
\label{fig:corners}
\end{figure}

\section{Results}\label{sec:results}

\subsection{Overview}

\begin{figure*}[!tb]
\centering
\includegraphics[width=0.9\textwidth]{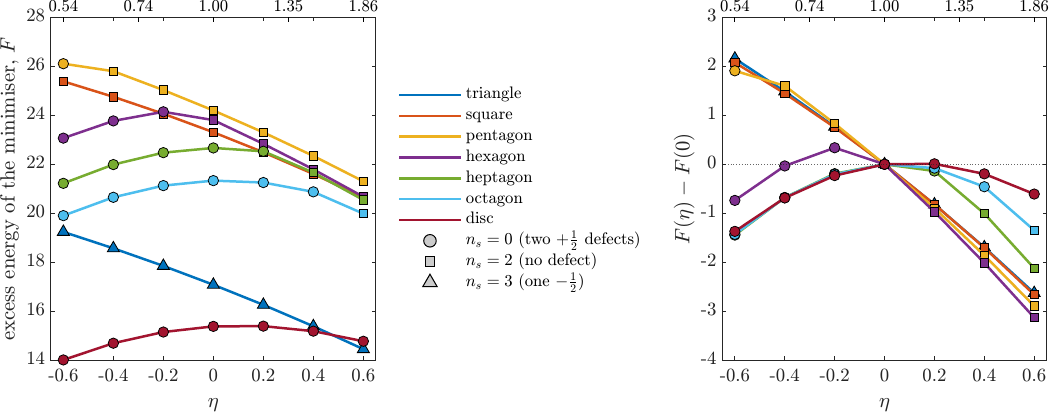}
\caption{Excess energy $F$ of the global minimiser, in units of $L_1s_+^2$, against the anisotropy $\eta$ (lower axis) and
$K_3/K_1$ (upper axis), for $\gamma=1$ and $\lambda=100$: (left) $F$; (right) $F(\eta)-F(0)$. The symbols give the number of splay
corners of the minimiser.}
\label{fig:minimisers}
\end{figure*}

\begin{figure*}[!tb]
\centering
\includegraphics[width=0.85\textwidth]{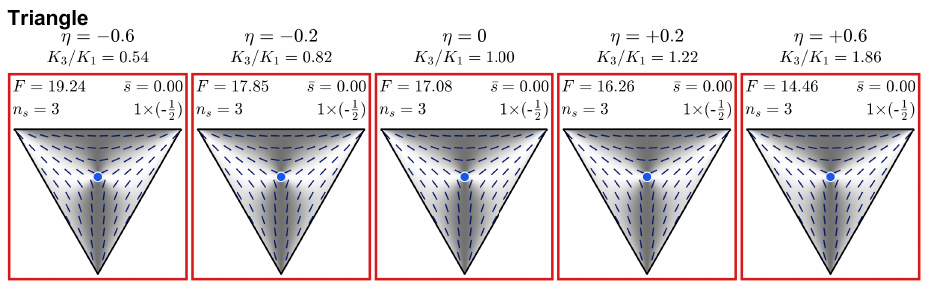}\\[4pt]
\includegraphics[width=0.85\textwidth]{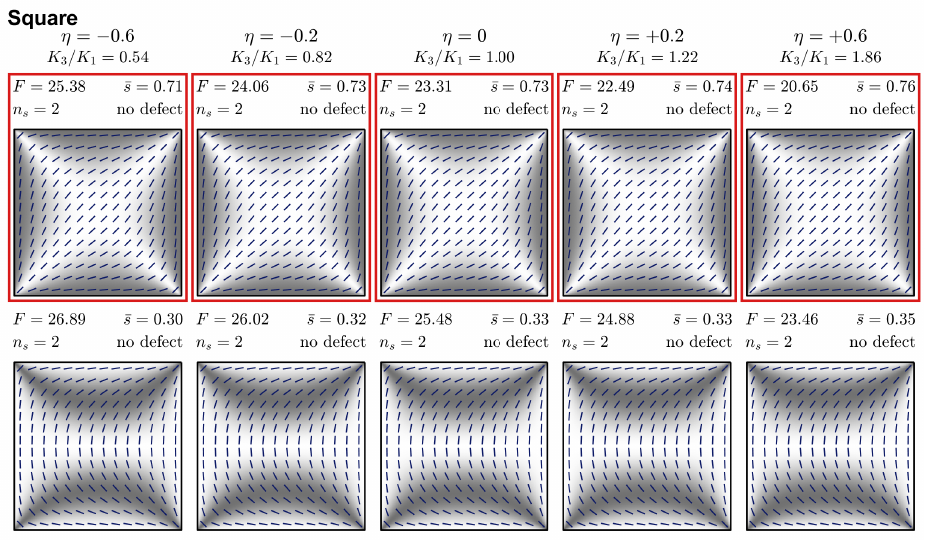}
\caption{Stable states of the triangle (top) and the square (bottom) for $\gamma=1$, $\lambda=100$. Columns: the anisotropy $\eta$, with
$K_3/K_1$; rows: distinct states, the global minimiser first. The grey scale is the intensity between crossed polarisers,
$I/I_0=(s/s_+)^2\sin^22(\varphi-\chi)$, with the polariser at $45^\circ$ to the average director, drawn from mid-grey ($I=0$) to
white ($I=I_0$) so that the director remains visible; the rods show the director. Defects of index $+\frac12$ are red and of
index $-\frac12$ blue; circles are interior defects and squares defects bound to a corner. Above each panel are the excess
energy $F$, in units of $L_1s_+^2$, the macroscopic order $\bar s$, the number of splay corners $n_s$, and the interior defects. A
red frame marks the global minimiser; an empty place means that the state is not stable at that $\eta$.}
\label{fig:tri_sq}
\end{figure*}

\begin{figure*}[!tb]
\centering
\includegraphics[width=0.85\textwidth]{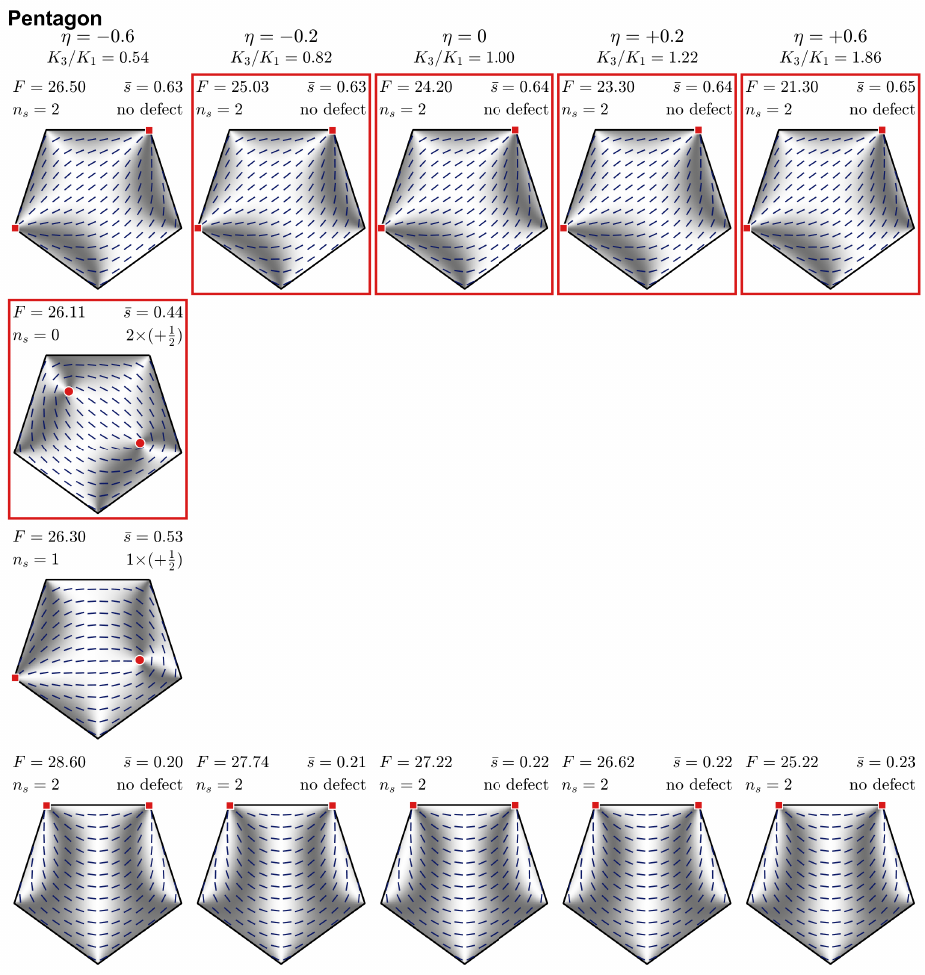}
\caption{Stable states of the pentagon, drawn as in Fig.~\ref{fig:tri_sq}. At $\eta=-0.6$, just below $\eta_c=-0.56$, the states with
no, one and two splay corners coexist with nearly equal energies, and the minimiser holds two interior defects of index
$+\frac12$; for $\eta\ge-0.2$ only states with two splay corners are stable.}
\label{fig:pentagon}
\end{figure*}

\begin{figure*}[!tb]
\centering
\includegraphics[width=0.85\textwidth,height=0.8\textheight,keepaspectratio]{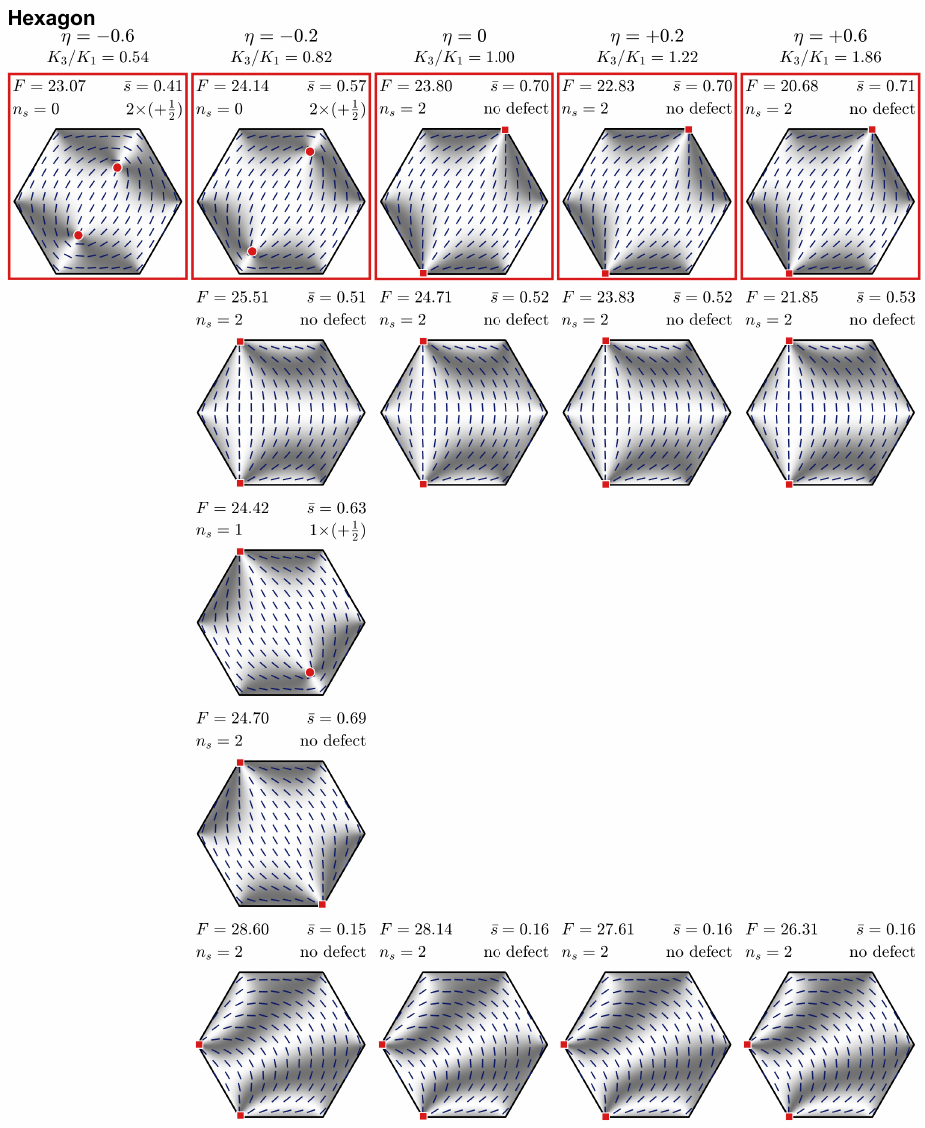}
\caption{Stable states of the hexagon, drawn as in Fig.~\ref{fig:tri_sq}. The interior $+\frac12$ defects of the minimiser move into
two opposite corners between $\eta=-0.2$ and $0$, as predicted by $\eta_c=0$. At $\eta=-0.2$ five states coexist; for $\eta\ge0$ the three
arrangements of two splay corners (opposite, one apart, adjacent) are stable.}
\label{fig:hexagon}
\end{figure*}

\begin{figure*}[!tb]
\centering
\includegraphics[width=0.85\textwidth,height=0.8\textheight,keepaspectratio]{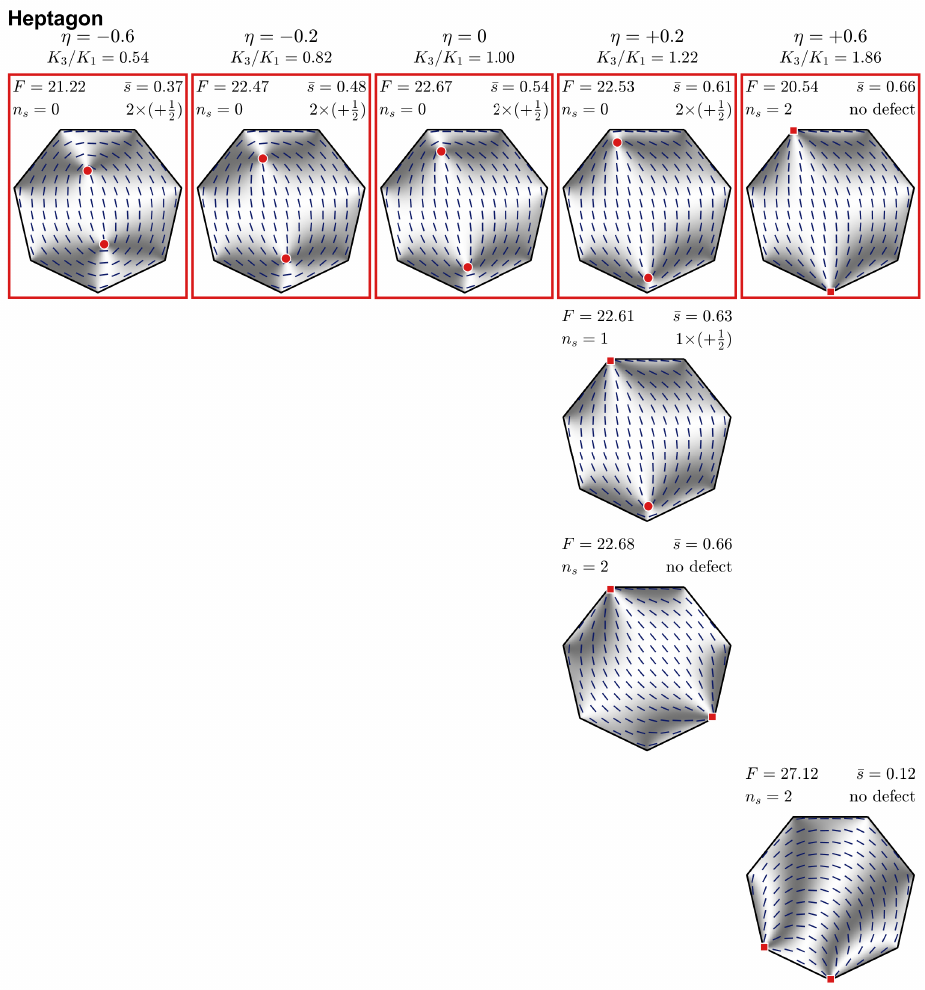}
\caption{Stable states of the heptagon, drawn as in Fig.~\ref{fig:tri_sq}. The interior $+\frac12$ defects approach the vertices as
$\eta$ increases and are absorbed between $\eta=0.2$ and $0.6$ ($\eta_c=0.28$); near the exchange, at $\eta=0.2$, three states are nearly
degenerate.}
\label{fig:heptagon}
\end{figure*}

\begin{figure*}[!tb]
\centering
\includegraphics[width=0.85\textwidth]{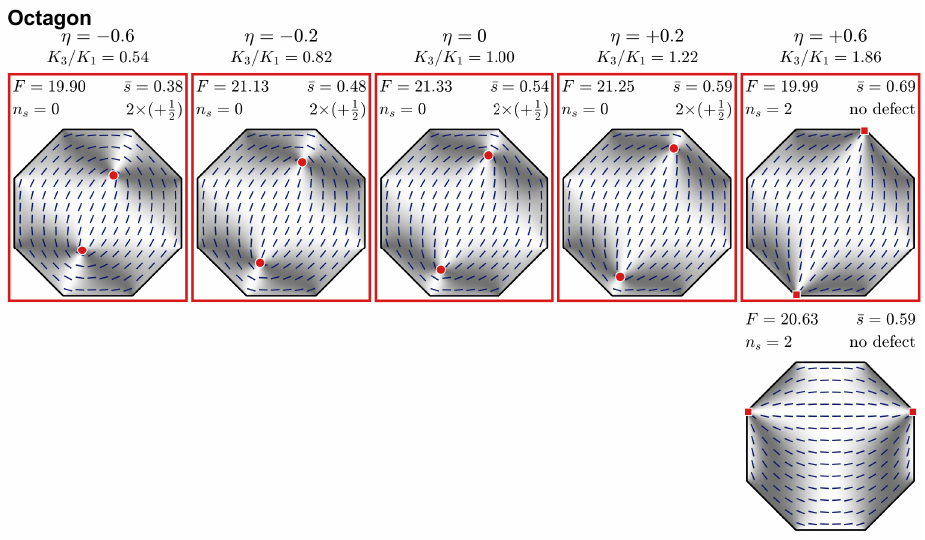}
\caption{Stable states of the octagon, drawn as in Fig.~\ref{fig:tri_sq}. The interior $+\frac12$ defects reach two opposite vertices
between $\eta=0.2$ and $0.6$ ($\eta_c=0.44$).}
\label{fig:octagon}
\end{figure*}

\begin{figure*}[!tb]
\centering
\includegraphics[width=0.85\textwidth]{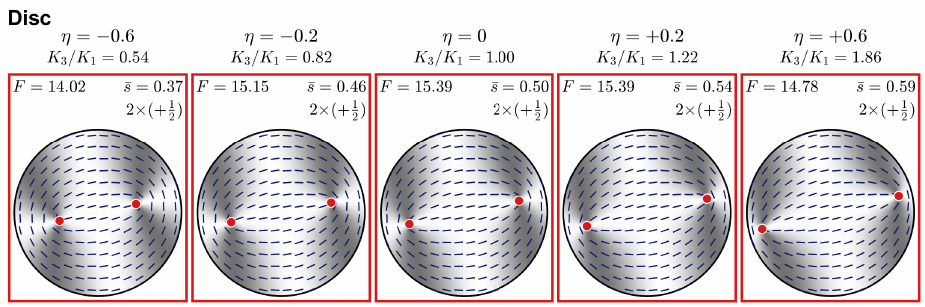}
\caption{The stable state of the disc, drawn as in Fig.~\ref{fig:tri_sq}. The two $+\frac12$ defects move towards the wall as $\eta$
increases.}
\label{fig:disc}
\end{figure*}

All the computations are for $\gamma=1$, $\lambda=100$ and tangential anchoring. Figs.~\ref{fig:tri_sq}, \ref{fig:pentagon}, \ref{fig:hexagon}, \ref{fig:heptagon}, \ref{fig:octagon} and \ref{fig:disc} show every
stable state found at $\eta=-0.6$, $-0.2$, $0$, $0.2$ and $0.6$, and Fig.~\ref{fig:minimisers} the energy of the global
minimiser at every computed $\eta$. The triangle and the disc have a single stable state over the whole range, and the
square two. The other polygons have up to five, the number being largest near the anisotropy $\eta_c$ of \eqref{eq:etac}:
four in the pentagon at $\eta=-0.6$, five in the hexagon at $\eta=-0.2$, three in the heptagon at $\eta=0.2$, two in the octagon
at $\eta=0.6$.

Fig.~\ref{fig:corners} compares the minimisers with the rule of \S\ref{sec:shed}. It plots the critical angles
$\omega_\pm(\eta)$ and $\omega_=(\eta)$, the interior angles of the polygons, and, for each polygon and each $\eta$, whether the
minimiser keeps splay corners (filled symbols) or holds two interior defects of index $+\frac12$ instead (open symbols).
Every one of the thirty minimisers lies on the side of the curve $\omega_+(\eta)$ that the rule predicts; the hexagon at
$\eta=0$, which lies on the curve, keeps its splay corners.

The energies of Fig.~\ref{fig:minimisers} follow the leading-order theory in their form, though not in their magnitude.
By lemma~\ref{lem:log}, only the splay corners contribute a term linear in $\eta$, equal to $-\frac12\eta$ times their total angle
times the logarithm. The minimisers of the triangle and the square, whose splay corners have the same total angle,
$3\times60^\circ=2\times90^\circ=180^\circ$, should therefore have the same slope, and they do: between $\eta=0$ and $0.2$ both
energies fall at the rate $dF/d\eta=-4.1$. The pentagon and the hexagon, in their states with two splay corners, have larger
total angles, $216^\circ$ and $240^\circ$, and fall faster, at $-4.5$ and $-4.9$, though not faster in proportion. A slope of
$-4.1$ for a total angle $\pi$ corresponds to an effective logarithm of about $2.6$, little more than half of $\ln\lambda\approx4.6$: at $\lambda=100$ the logarithm of a corner is
cut off well within the domain, by its neighbours, and the energies of order one are substantial. The states whose
interior holds two $+\frac12$ defects and no splay corners---the disc, and the heptagon and octagon for $\eta\le0.2$---have no
term linear in $\eta$ at leading order, and their energies are indeed nearly flat near $\eta=0$ and fall on both sides, as the
bound of lemma~\ref{lem:log}(c) requires.

\subsection{The triangle and the square}

The triangle (Fig.~\ref{fig:tri_sq}, top) has one stable state for every $\eta$: three splay corners and a defect of index
$-\frac12$ at the centroid, with $\bar s=0$ by symmetry. Its angle lies below $\omega_-$ throughout, so that, as \S\ref{sec:polygons}
predicts, no half-turn corner, and so no other state, survives. The energy falls from $19.24$ at $\eta=-0.6$ to $14.46$ at $\eta=0.6$,
because all the turning of the boundary is taken up by splay corners, which become cheaper as $\eta$ increases.

The square (Fig.~\ref{fig:tri_sq}, bottom) has the two states known in the one-constant model~\cite{Tsakonas07,LMR12}, the
diagonal state, with its splay corners at the ends of a diagonal, and the rotated state, with its splay corners at the
ends of an edge. Both have $n_s=2$ and no defects for every $\eta$, as \eqref{eq:cn} predicts. The diagonal state is the
minimiser throughout, with $\bar s$ between $0.71$ and $0.76$; the rotated state, with $\bar s\approx0.3$, lies above it by an
amount that grows from $1.51$ at $\eta=-0.6$ to $2.81$ at $\eta=0.6$. The two have the same corners, so that their energies differ
only at order one, and the anisotropy does not change their order.

\subsection{The pentagon to the octagon}

These polygons have angles between $108^\circ$ and $135^\circ$, close to $\omega_+$, and in each the minimiser changes, within the range
of the computations, between a state with two splay corners and a state with two interior defects of index $+\frac12$. The
change occurs within the interval of $\eta$ predicted both by \eqref{eq:etac} and by the exact values of
\S\ref{sec:exact}: between $-0.6$ and $-0.2$ in the pentagon ($\eta_c=-0.56$, exactly $-0.45$), between $-0.2$ and $0$ in the
hexagon ($\eta_c=0$), and between $0.2$ and $0.6$ in the heptagon and the octagon ($\eta_c=0.28$ and $0.44$, exactly $0.32$ and
$0.57$). The computations, at intervals of $0.2$ in $\eta$, bracket each exchange but do not resolve the difference between
the first-order and exact critical anisotropies, which is at most $0.13$; nor would they separate it from the shift due to
the energies of order one, which at $\lambda=100$ may be of the same size.

The mechanism is the one described in \S\ref{sec:shed}. In the hexagon (Fig.~\ref{fig:hexagon}) the minimiser at $\eta=-0.6$
holds two $+\frac12$ defects in the interior, near opposite corners. As $\eta$ increases they move towards those corners, and
between $\eta=-0.2$ and $0$ they are absorbed, turning the two half-turn corners into splay corners; the branch continues as the
state with splay corners at opposite vertices, which is the minimiser for $\eta\ge0$. The same migration is seen in the heptagon
and the octagon (Fig.~\ref{fig:heptagon} and Fig.~\ref{fig:octagon}), where the defects reach the vertices between $\eta=0.2$ and $0.6$.
In the pentagon (Fig.~\ref{fig:pentagon}), whose angle is the smallest of the four, the states with interior defects are
stable only at $\eta=-0.6$, the one value computed below its $\eta_c$. The transition is thus continuous along a branch: a defect that approaches a corner near the critical angle is
the hidden defect of \S\ref{sec:shed}, held near the vertex by the energies of order one.

Near the exchange the states with no, one and two splay corners are nearly degenerate, as the linearity of \eqref{eq:cn}
in $n_s$ requires, and they coexist as local minimisers. In the pentagon at $\eta=-0.6$ their energies are $26.11$, $26.30$ and
$26.50$; in the hexagon at $\eta=-0.2$ they are $24.14$, $24.42$ and $24.70$. In both cases the three are equally spaced, and
ordered with $n_s=0$ lowest, as they should be for $\eta$ just below $\eta_c$. The spacing, of order a few tenths, is much less
than the energies themselves, and this near-degeneracy is the source of the multistability of these polygons. In the
heptagon at $\eta=0.2$, just below $\eta_c=0.28$, three states lie within $0.15$ of each other.

The arrangement of the splay corners matters at order one. In the pentagon, the state with splay corners at two
non-adjacent vertices lies below the state with adjacent splay corners by $2.1$ at $\eta=-0.6$ and $3.9$ at $\eta=0.6$. In the
hexagon, with splay corners at opposite vertices, at vertices one apart, and at adjacent vertices, the energies at $\eta=0$
are $23.80$, $24.71$ and $28.14$. As in the square, splay corners prefer to be far apart, and the preference grows with $\eta$.

\subsection{The disc}

The disc (Fig.~\ref{fig:disc}) has a single stable state for every $\eta$, with two defects of index $+\frac12$ on a diameter,
as predicted for $|\eta|<1$. There are no corners to absorb them, but they move towards the wall as $\eta$ increases, from about
$0.4R$ from the centre at $\eta=-0.6$ to about $0.9R$ at $\eta=0.6$, and $\bar s$ increases from $0.37$ to $0.59$. This is the
continuous counterpart of the exchange in the polygons: by \S\ref{sec:polygons}, in the limit of many sides the two defects
become two splay corners at $\eta=1$. The energy is greatest near $\eta=0.1$ and falls on both sides, by $1.37$ at $\eta=-0.6$ and
$0.61$ at $\eta=0.6$. We attribute the asymmetry to the bend of the director that follows the curved wall, which is of order one
and costs more when $\eta>0$.

\subsection{The bound on the order}\label{sec:boundres}

The bound \eqref{eq:bound} is never active. In every domain, at every $\eta$ computed, and in every one of the 82 stable
states found, there is no node at which the order reaches $\frac12$ with the energy decreasing as it increases; the
multiplier $\mu$ of \eqref{eq:EL} therefore vanishes everywhere, and every state shown is a classical solution of the
Euler--Lagrange equations \eqref{eq:EL}--\eqref{eq:BC}. This is as \S\ref{sec:rise} predicts, since $|\eta|\le0.6$ lies well within
\eqref{eq:43}. The order nonetheless comes close to the bound: its largest value falls short of $\frac12$ by between
$8\times10^{-8}$, in the octagon, and $4\times10^{-7}$, in the triangle, and in every domain it is largest at the greatest
anisotropy computed, $\eta=0.6$, as the estimate $1-\tilde s\approx(2-\frac32\eta\zeta)G/\lambda^2$ of \S\ref{sec:rise} leads one to expect
in the weakly distorted, splay-dominated regions of the splay-corner states.

To test the threshold \eqref{eq:43} itself we computed the square for $1.2\le|\eta|\le1.45$, in steps of $0.05$, at $\lambda=20$ and
$30$ and with $\gamma=1$ and $0.3$. In all four cases the bound is inactive in every stable state for $|\eta|\le1.3$, and is
active in the minimiser at $\eta=1.35$, and in some of the stable states at $\eta=-1.35$; the threshold thus lies between $1.30$
and $1.35$, in agreement with $\frac43\approx1.333$, and shows no dependence on $\lambda$ or $\gamma$ over the values tested, as
\S\ref{sec:rise} predicts. For
$\eta>\frac43$ the active nodes lie in the regions of nearly pure splay that spread from the splay corners, and their number grows
rapidly with $\eta$; at $\eta=1.5$, $\lambda=30$ and $\gamma=1$ they cover about a quarter of the domain in the minimiser. For
$\eta<-\frac43$ they lie in star-shaped regions of nearly pure bend, and further stable states appear, among them states with
interior defects, which exist only while the bound is active and merge with the familiar states when $|\eta|$ is reduced
below $\frac43$. Beyond the threshold the bound no longer serves only to make the problem well posed but shapes the
equilibria; this is the regime discussed at the end of the Conclusions.

\section*{Conclusions}

In a confined nematic large compared with its coherence length, the corners of the container are sites at which defects
are exchanged with the liquid. A corner of opening angle $\omega$ can take several types, distinguished by the number of
half-turns of the director relative to the radius; neighbouring types differ by a bound defect of index $\pm\frac12$; and a
balance of logarithmic energies gives the force on a defect near the vertex, and so the critical angles at which the corner
sheds or absorbs it, $72^\circ$, $120^\circ$, $216^\circ$ and $360^\circ$ in a one-constant nematic, with windows of bistability between
them. The hidden defect observed near a $72^\circ$ vertex, and the limits of the trefoil state in isosceles triangles, are
instances of this exchange. Applied corner by corner, the rule selects the equilibria of regular polygons: the triangle,
whose angles lie below $72^\circ$, has only splay corners and a defect of index $-\frac12$; the square lies inside the window of
bistability; and the polygons whose angles are near $120^\circ$ choose between splay corners and interior defects of index
$+\frac12$. The same local rule selects the corner types of any convex polygon: the two sharpest corners are splay corners if
their angles lie below $\omega_+$, and every further corner only if its angle lies below $\omega_-$. For an isosceles triangle this
places the trefoil state between apex angles of $36^\circ$ and $72^\circ$, as found numerically.

Splay-bend anisotropy acts on this picture almost entirely through the splay corner. In two dimensions it can be
represented only by a term cubic in the order tensor, and that term makes the Landau--de Gennes energy unbounded below; the
requirement that the order not exceed its bulk value removes the difficulty without any addition to the energy, for exactly
those anisotropies at which both Frank constants are positive. The bound is, moreover, expected to be inert for every known
nematic: away from the defect cores the order can rise above its bulk value only where the distortion is nearly a pure splay
or a pure bend and $|\eta|>\frac43$, so that for $\frac15<K_3/K_1<5$ the bound is predicted to remain inactive outside the cores. The
computations are consistent with this: they place the threshold between $|\eta|=1.30$ and $1.35$, with no dependence on the
size of the domain or the anchoring strength over the values tested, and find no active bound below it. The idea is simple
and physically natural: the cubic invariant becomes unbounded only when it is asked to act on an order larger than the
nematic has in bulk, and bounding the order by its bulk value, a property the one-constant equilibria possess in any case,
is enough to make the theory well posed. The same idea, a bound on the magnitude of the order tensor relative to its bulk
value, can be carried over to the full Landau--de Gennes theory in three dimensions, where the order tensor has five
components and the quadratic invariants contribute their own anisotropy; this extension will be presented in a forthcoming
paper. To first order only the splay corner feels the anisotropy; the energies of the
other corners and of half-integer defects are even in it, and follow exactly from a quadrature. The angle above which a splay
corner sheds a $+\frac12$ defect moves from $104^\circ$ to $136^\circ$ as $K_3/K_1$ increases from $0.54$ to $1.86$, and for a regular
polygon this gives, to first order, the anisotropy $\eta_c=N(N-6)/(N-2)^2$ at which states with two splay corners exchange
with states holding two interior $+\frac12$ defects. The computations confirm the rule: in every polygon the minimiser lies on
the side of it that is predicted, the exchange takes place by the migration of the defects into the corners, and near it
several states coexist with nearly equal energies. Since homeotropic anchoring is tangential anchoring with the sign of the
anisotropy reversed, the same rules apply to homeotropic walls with splay and bend exchanged.

Four questions remain open. The computations test the critical angle near $120^\circ$ and its movement with the anisotropy;
the angles $216^\circ$ and $360^\circ$ of re-entrant corners, and the bistability between them, could be tested in star-shaped or
L-shaped wells. At $\lambda=100$ the effective logarithm is little more than half of $\ln\lambda$, so that the energies of order one
shift the exchanges by amounts that may be comparable with the higher-order corrections of \S\ref{sec:exact}; a computation
at larger $\lambda$, or a calculation of the order-one terms, would separate the two. The rule has been derived with
strong anchoring; with weaker anchoring a defect held at a vertex merges with the surface defect of the corner, and the
critical angles should then depend on the anchoring strength.

Finally, a nematic with an anisotropy far larger than that of any known material, $|\eta|>\frac43$, might one day be
discovered or synthesised. The constrained theory remains well posed for it, but describes it by the Oseen--Frank energy
alone; the bound, moreover, then becomes active over large parts of the domain and creates states of its own
(\S\ref{sec:boundres}), and a material so anisotropic may well show distortions of a kind that this energy does not contain. The singular
potential of Ball and Majumdar~\cite{BM10} keeps the eigenvalues of the order tensor within their microscopic limits. Since
the tensor of Landau--de Gennes theory is a renormalised, coarse-grained quantity (\S\ref{sec:order}), we do not regard that
bound as the natural one for it; imposing the microscopic limit on the coarse-grained tensor may also produce thin layers,
near defects and walls, in which the order changes sharply, without a counterpart in the material. For such a material the
quartic elastic terms of Golovaty, Novack and Sternberg~\cite{GNS21} seem the more appropriate route, for they introduce new
elastic constants, and a nematic of such strong anisotropy would be expected to reveal them through distortions of a new
kind. For the known nematics, whose anisotropies lie well within $|\eta|<2$, those additional constants would not be fixed
by any measurement, and the bound on the order is all that is needed.

\section*{Appendix}

\appendix
\section{Proof of theorem~\ref{thm:main}}\label{app:proof}

\emph{(i)} For $\q\in\Kq_{\bar q}$ the eigenvalues of $\A(\q)$ lie between $c_0=1-|\eta|\bar q>0$ and $2-c_0$, so that
\begin{equation}\label{eq:coerc}
c_0|\nabla\q|^2\le\sum_a\nabla q_a\cdot\A(\q)\nabla q_a\le(2-c_0)|\nabla\q|^2 .
\end{equation}
Each term of $F$ is nonnegative, and $F[\mathbf 0]$ is finite. Let $(\q_n)$ be a minimising sequence, that is, a sequence in
$\Kq_{\bar q}$ with $F[\q_n]\to\inf F$. By \eqref{eq:coerc}, $\int|\nabla\q_n|^2dA\le F[\q_n]/c_0$, and $\int|\q_n|^2dA\le\bar q^2|\Omega|$,
so that the sequence is bounded in $H^1$. By the Rellich--Kondrachov theorem a subsequence converges weakly in $H^1$, strongly
in $L^2(\Omega)$ and almost everywhere, and its boundary values converge in $L^2(\partial\Omega)$, the trace operator being compact on
bounded Lipschitz domains~\cite{Necas}. The set $\Kq_{\bar q}$ is convex and closed in $H^1$, hence closed under weak
convergence, and contains the limit $\q$. Let $P_{\bar q}$ be the radial projection onto the disc of radius $\bar q$, defined as in
\eqref{eq:proj} with $\frac12$ replaced by $\bar q$. The function $f(\mathbf u,\mathbf p)=\sum_a\mathbf p_a\cdot\A(P_{\bar q}(\mathbf
u))\mathbf p_a$ is continuous, nonnegative, convex in the gradient variable $\mathbf p$, and coincides with the elastic density
on $\Kq_{\bar q}$, so that its integral is weakly lower semicontinuous~\cite[Thm.~3.23]{Dacorogna}. The Landau density is
continuous and bounded on $\Kq_{\bar q}$, so that its integral converges by dominated convergence, and the anchoring integral
converges with the boundary values. Hence $F[\q]\le\liminf F[\q_n]$, and $\q$ is a minimiser.

\emph{(ii)} Choose $r$ with $1/|\eta|<r<\bar q$, and $\delta>0$ with $r+\delta\le\bar q$ and $|\eta|\delta\le\frac12(|\eta|r-1)$. Let $\q_0$ be a
constant with $|\q_0|=r$, $\mathbf e$ a unit eigenvector of $\A(\q_0)$ for its eigenvalue $1-|\eta|r<0$, $\mathbf v$ a unit vector,
and $\q_k=\q_0+\delta\sin(k\,\mathbf e\cdot\mathbf x)\,\mathbf v$. Then $|\q_k|\le\bar q$, and, since the eigenvalues of
$\A(\q_k)-\A(\q_0)$ are at most $|\eta|\delta$ in magnitude, the elastic density is at most
$-\frac12(|\eta|r-1)\delta^2k^2\cos^2(k\,\mathbf e\cdot\mathbf x)$, whose integral tends to $-\infty$ like $-k^2$. The Landau and anchoring
energies are bounded independently of $k$.

\paragraph{The variational inequality.} A minimiser $\q$ satisfies, for every $\mathbf v\in\Kq_{\bar q}$ and $\mathbf w=\mathbf v-\q$,
\begin{multline}
\int_\Omega\Big[2\sum_a\nabla w_a\cdot\A(\q)\nabla q_a+\eta\,\mathbf w\cdot\mathbf S(\nabla\q)\\
+8\lambda^2\big(|\q|^2-\tfrac14\big)\q\cdot\mathbf w\Big]dA\\
+2\gamma\lambda\oint_{\partial\Omega}(\q-\q_b)\cdot\mathbf w\,ds\ge0,
\end{multline}
where $\mathbf S(\nabla\q)=\big(|\partial_x\q|^2-|\partial_y\q|^2,\ 2\,\partial_x\q\cdot\partial_y\q\big)$ is the vector of
\eqref{eq:EL}, the second term being integrable because $\mathbf w$ is bounded. On an open set where $|\q|\le\bar q-\epsilon$ for some $\epsilon>0$,
perturbations of both signs are admissible and the inequality becomes the weak form of \eqref{eq:EL}--\eqref{eq:BC} with
$\mu=0$.

\section{The discrete cubic term}\label{app:discrete}

On a triangle $T$ with vertices $v=1,2,3$, let $\beta^x_v$ and $\beta^y_v$ be the derivatives of the linear basis functions, and
$\q_v$ the nodal values. The gradient, $\mathbf g^x=\sum_v\beta^x_v\q_v$ and $\mathbf g^y=\sum_v\beta^y_v\q_v$, is constant, and the
value at the centroid is $\mathbf p=\frac13\sum_v\q_v$. The cubic energy of the triangle is exactly
\begin{gather}
E_T=\eta|T|\,c(\mathbf p,\mathbf g^x,\mathbf g^y),\nonumber\\
c=p_1\big(|\mathbf g^x|^2-|\mathbf g^y|^2\big)+2p_2\,\mathbf g^x\cdot\mathbf g^y .
\end{gather}
Its derivative with respect to the component $a$ of $\q_v$ is
\begin{equation}
\frac{\partial E_T}{\partial q_{v,a}}=\eta|T|\Big[\frac13\frac{\partial c}{\partial p_a}+\beta^x_v\frac{\partial c}{\partial g^x_a}+\beta^y_v\frac{\partial c}{\partial g^y_a}\Big],
\end{equation}
with $\partial c/\partial\mathbf p=(|\mathbf g^x|^2-|\mathbf g^y|^2,\ 2\mathbf g^x\cdot\mathbf g^y)$,
$\partial c/\partial\mathbf g^x=2p_1\mathbf g^x+2p_2\mathbf g^y$ and $\partial c/\partial\mathbf g^y=-2p_1\mathbf g^y+2p_2\mathbf g^x$. Since $c$ is cubic and
$(\mathbf p,\mathbf g^x,\mathbf g^y)$ linear in the nodal values, the Hessian is
\begin{multline}
\frac{\partial^2E_T}{\partial q_{v,a}\partial q_{w,b}}=\eta|T|\Big\{\delta_{ab}\big[2p_1(\beta^x_v\beta^x_w-\beta^y_v\beta^y_w)\\
+2p_2(\beta^x_v\beta^y_w+\beta^y_v\beta^x_w)\big]\\
+\tfrac13\big[U^x_{ab}\beta^x_w+U^y_{ab}\beta^y_w\big]
+\tfrac13\big[U^x_{ba}\beta^x_v+U^y_{ba}\beta^y_v\big]\Big\},
\end{multline}
where $U^x_{ab}=\partial^2c/\partial p_a\partial g^x_b$ and $U^y_{ab}=\partial^2c/\partial p_a\partial g^y_b$, that is,
$U^x_{1b}=2g^x_b$, $U^y_{1b}=-2g^y_b$, $U^x_{2b}=2g^y_b$ and $U^y_{2b}=2g^x_b$. These expressions were checked against finite
differences, and the discrete energy of a slowly varying director field against \eqref{eq:OF}.

\section{The constrained Newton iteration}\label{app:newton}

The nodal constraints are $c_i=|\q_i|^2-\bar q^2\le0$. At an iterate $\q$ with gradient $\mathbf g$ of the discrete energy, a node
is active if $|\q_i|=\bar q$ and $\mathbf g_i\cdot\q_i<0$, so that the energy would decrease by increasing the order there; its
multiplier is $\mu_i=-\mathbf g_i\cdot\q_i/(2|\q_i|^2)\ge0$. Let $Z$ be the matrix whose columns span the permitted directions,
both components of $\q$ at the free nodes and the tangential direction $(-q_{i2},q_{i1})/|\q_i|$ at the active nodes, and let
$H_L=H+\operatorname{diag}(2\mu_i)$ be the Hessian of the Lagrangian, $H$ being the Hessian of the energy. The step is
$\delta\q=Z\,\delta\mathbf z$, with
\begin{equation}
\big(Z^{\mathsf T}H_LZ+\kappa Z^{\mathsf T}DZ\big)\,\delta\mathbf z=-Z^{\mathsf T}\mathbf g,
\end{equation}
$D$ being the lumped mass matrix scaled by $\lambda^2$ and $\kappa$ a damping parameter, adjusted so that the energy decreases
monotonically; after the step the nodes outside the bound are projected onto it. An equilibrium is stable if
$Z^{\mathsf T}(H_L-\epsilon M)Z$ is positive definite, $M$ being the mass matrix and $\epsilon=10^{-6}$; when no node is active this is
the usual test on the Hessian.

\balance
\bibliographystyle{unsrtnat}
\bibliography{anisotropic_rldg}

\end{document}